\documentclass[
aps,prd,
12pt,
nopreprintnumbers,
showpacs,
eqsecnum,
nofootinbib
]{revtex4-1}

\usepackage{graphicx}
\usepackage{amssymb}

\begin{document}

\title{Integrable higher-dimensional cosmology with separable variables in
an Einstein-dilaton-antisymmetric f{}ield theory}
\author{Nahomi Kan}\email[]{kan@gifu-nct.ac.jp}
\affiliation{National Institute of Technology, Gifu College,
Motosu-shi, Gifu 501-0495, Japan
}
\author{Masashi Kuniyasu}\email[]{mkuni13@yamaguchi-u.ac.jp}
\author{Kiyoshi Shiraishi}\email[]{shiraish@yamaguchi-u.ac.jp}
\author{Kohjiroh Takimoto}\email[]{i016vb@yamaguchi-u.ac.jp}
\affiliation{
Graduate School of Sciences and Technology for Innovation, Yamaguchi
University, Yamaguchi-shi, Yamaguchi 753--8512, Japan}
\date{\today}

\begin{abstract}
We consider a $D$-dimensional cosmological model with a dilaton field and two
$(D-d-1)$-form field strengths which have nonvanishing fluxes in extra dimensions. 
Exact solutions for the model with a certain set of couplings are
obtained by separation of three variables. Some of the solutions describe
accelerating expansion of the $d$-dimensional space.
Quantum cosmological aspects of the model are also briefly mentioned.
\end{abstract}



\maketitle

\section{Introduction}
\label{sec1}
In recent decades,
relativistic models with a scalar field have received much
interest in cosmology because they are suitable for the inflationary scenario
\cite{inflation} and for tackling the dark energy problems
\cite{darkenergy1,darkenergy2}. Although numerical solutions or approximate
solutions for scale factors are studied in many cosmological models,
it would be very interesting to find exact solutions that describe
the accelerating universe.
In many areas of physics, exact solutions play the most important role
in understanding and growing the crude concepts.
Recently, many authors have studied integrable models with exponential scalar
potentials for cosmology and found various interesting ones; for example,
some models account for the transient acceleration of the universe
\cite{STT,Russo,ENO,FSS,FST,SS,MP,PM,ANL,ALNW}.

On the other hand, it is known that
dilaton gravity arises from a low-energy effective theory of string theory,
from certain supergravity theories, 
and from higher-dimensional theories with extra dimensions.
In models based on such theories, scalar fields naturally appear with exponential
potentials.
In addition, such theories often contain totally antisymmetric tensor
fields; the configuration on the extra space can play an important role in
compactification scheme \cite{MN1,MN2,GT,Halliwell1,Halliwell2}. Some
higher-dimensional integrable cosmological models  have been studied
\cite{Sano,TW,Ohta2} and exact analytic solutions have been derived in some
specific cases with fluxes in extra dimensions 
\cite{LMPX,Kaloper,CGG,Roy,Ohta1,Ohta3,CHNOW,Ohta4}.

In the present paper, we consider analytically solvable models of
$D$-dimensional cosmology with a scalar dilaton and antisymmetric
tensor fields.
Integrability does not necessarily mean the existence of analytic solutions.
We propose a model in which the equations of motion
can be expressed by three separate equations of Liouville type;
then, its cosmological solutions can be written in simple elementary functions.
We
analyze the simple model on the basis of the possibility of temporal accelerating
expansion of the $(d+1)$-dimensional universe ($d+1<D$).

Here, we illustrate an essential structure of the solvable model we consider in
this paper. Provided that the classical cosmological action (in the
minisuperspace) can be written in the form
\begin{equation}
S=\int
dt\sum_a\left[\sigma_a\frac{1}{2}\dot{x}_a^2-\frac{V_a}{2}
e^{2\lambda_ax_a}\right]\,,
\end{equation}
where $\sigma_a=\pm 1$ and the dot denotes the derivative with respect to $t$,
the equations of motion give the one-dimensional Liouville equation
\begin{equation}
\sigma_a\ddot{x}_a+\lambda_aV_a
e^{2\lambda_ax_a}=0\,.
\end{equation}
Then, the integrals of motion are 
\begin{equation}
E_a\equiv\sigma_a\frac{1}{2}\dot{x}_a^2+\frac{V_a}{2}
e^{2\lambda_ax_a}\,.
\end{equation}

We obtain the following analytic solutions for the equations of motion:
\begin{itemize}
\item For $\sigma_aV_a>0$,
\begin{equation}
x_a(t)=\frac{1}{2\lambda_a}\ln\frac{q_a^2}{\cosh^2q_a\sqrt{|V_a|}\lambda_a
(t-t_a)}\,,
\quad E_a=\sigma_a\frac{q_a^2|V_a|}{2}\,.
\end{equation}
\item For $\sigma_aV_a<0$,
\begin{equation}
x_a(t)=\frac{1}{2\lambda_a}\ln\frac{q_a^2}{\sinh^2q_a\sqrt{|V_a|}\lambda_a
(t-t_a)}\,,
\quad E_a=\sigma_a\frac{q_a^2|V_a|}{2}\,,
\end{equation}
\begin{equation}
x_a(t)=\frac{1}{2\lambda_a}\ln\frac{q_a^2}{\sin^2q_a\sqrt{|V_a|}\lambda_a
(t-t_a)}\,,
\quad E_a=-\sigma_a\frac{q_a^2|V_a|}{2}\,.
\end{equation}
\item For $V_a=0$, we find that $x_a(t)= q_a(t-t_a)$ and
$E_a=\sigma_a\frac{q_a^2}{2}$,
\end{itemize}
where $q_a$ and $t_a$ are integration constants.

Remembering that the model is a cosmological one, the Hamiltonian constraint
restricts the constants as
\begin{equation}
\sum_a\left[\sigma_a\frac{1}{2}\dot{x}_a^2+\frac{V_a}{2}
e^{2\lambda_ax_a}\right]=\sum_a E_a=0\,.
\end{equation}
We present a model that is soluble by using such a separation
of variables in the next section.
Furthermore, we will see later that the separation of variables is significant for
considering the minisuperspace Wheeler--De\,Witt equation.

The outline of the present paper is as follows.
In Sec.~\ref{av}, we define our models in which three variables are separable as
in the manner mentioned above.
The solutions are exhibited in Secs.~\ref{cano} and \ref{phan}, and summarized in
Appendix~\ref{ss}. Section~\ref{cano} is devoted to the solutions for the scalar
field with the canonical kinetic term, while Sec.~\ref{phan} treats the case of the
``phantom'' scalar field. The physical scale factor and the physical property of
the solutions are discussed in Sec.~\ref{ac}. Section~\ref{qc} contains a
brief description of the quantum cosmology of our model through the minisuperspace
Wheeler--De\,Witt equation. We conclude with a discussion in Sec.~\ref{sd}.
\section{action and variables}
\label{av}
Let us consider the action of the $D$-dimensional model
\begin{equation}
S=\int
d^Dx\sqrt{-g}\left[R-\sigma\frac{1}{2}(\nabla\Phi)^2-
\frac{l}{2p!}
e^{2\kappa\alpha\Phi}{F_{[p]}^{(l)}}^2-\frac{r}{2p!}
e^{-2\kappa\frac{\sigma}{\alpha}\Phi}{F_{[p]}^{(r)}}^2
\right]\,,
\label{action}
\end{equation}
where  $R$ is the Ricci scalar derived from the metric $g_{MN}$
($M,N=0,1,\dots,D-1$), $g$ is the determinant of $g_{MN}$, and $\Phi$ is a real
scalar field which has dilatonlike coupling to the two $p$-form field strengths
$F_{[p]}^{(l)}$ and $F_{[p]}^{(r)}$. The constant $\alpha$ represents a scalar
(dilaton) self coupling constant. The constants $l$ and $r$ indicate the
couplings between the scalar and the two antisymmetric tensor field strengths. We
also use the abbreviations
$(\nabla\Phi)^2\equiv
g^{MN}\partial_M\Phi\partial_N\Phi$ and
$F^2_{[p]}=g^{M_1N_1}g^{M_2N_2}\cdots g^{M_{p}N_{p}}
{F_{[p]}}_{M_1M_2\cdots M_p}{F_{[p]}}_{N_1N_2\cdots N_p}$.

The action (\ref{action}) is invariant under the following two independent
transformations:
\begin{eqnarray}
(1)& &\quad\alpha\leftrightarrow -\alpha\quad\mbox{and}\quad \Phi\leftrightarrow
-\Phi\,,\\ (2)& &\quad\alpha\leftrightarrow -\sigma/\alpha\quad\mbox{and}\quad
l\leftrightarrow r\quad\mbox{and}\quad F_{[p]}^{(l)}\leftrightarrow
F_{[p]}^{(r)}\,.
\end{eqnarray}
If the constant $\sigma$ is taken to be $\sigma=+1$,
the kinetic term of the scalar field becomes a canonical one.
If we choose $\sigma=-1$, the scalar becomes a phantom field \cite{phantom}.
Due to the symmetries, we only have to investigate the cases with $0<\alpha\le 1$
to clarify the general behaviors of the system.%
\footnote{This is not the case for $l=0$ or $r=0$.}
Note also that the choice $r=0$ in the action (\ref{action}) reduces the model
to that studied by many other authors, such as in Refs.~\cite{CGG,Roy}.%
\footnote{The solutions for the model with $r=0$ are discussed in
Appendix~\ref{r0}.} 
The value of the constant $\kappa$ will be specified later.

We adopt the following ans\"atze.
We assume the $(D-1)$-dimensional space admits the metric of a direct product of a
$d$-dimensional flat Euclidean space and $(D-d-1)$-dimensional maximally symmetric
space.
The scale factors and the scalar $\Phi$ are considered to be only time-dependent;
i.e., they are functions of the time coordinate $t=x^0$. Therefore, we take the
metric as follows:
\begin{equation}
ds^2=g_{MN}dx^Mdx^N=-e^{2n(t)}dt^2+e^{2a(t)}d\mathbf{x}^2+
e^{2b(t)}d\Omega_{D-d-1}^2\,.
\end{equation}
Here, we denote the coordinates of the flat space as $x^i$ ($i=1,\dots,d$)
and those of the maximal symmetric extra space as $x^m$ ($m=d+1,\dots,D-1$).
We use the notation $d\mathbf{x}^2\equiv\sum_{i=1}^d (dx^i)^2$, and
$d\Omega_{[D-d-1]}^2$ stands for the line element of the extra space
whose metric is denoted as $\tilde{g}_{mn}$. 
We assume that the Ricci tensor of the extra space is written as
\begin{equation}
\tilde{R}_{mn}=k_b(D-d-2)\tilde{g}_{mn}\,,
\end{equation}
where $k_b$ is a constant, which has been normalized to $1$, $0$, or $-1$.

We further consistently assume that the $p$-form field strengths take 
``constant'' (flux) values in the extra space; thus, 
\begin{equation}
p=D-d-1\,
\end{equation}
and
\begin{equation}
{F_{[D-d-1]}^{(l)}}_{d+1,d+2,\cdots,{D-1}}=
{F_{[D-d-1]}^{(r)}}_{d+1,d+2,\cdots,{D-1}}=f\,,
\end{equation}
where $f$ is taken as a positive constant (and is possibly quantized as a
``magnetic'' charge), without loss of generality. Even if we assume 
nonidentical values for two fluxes as classical configurations, the
difference in the magnitudes of the fluxes
 can be absorbed into the redefinition of the couplings
$l$ and $r$. We now obtain
\begin{equation}
\frac{1}{(D-d-1)!}({F_{[D-d-1]}^{(l)}})^2=
\frac{1}{(D-d-1)!}({F_{[D-d-1]}^{(r)}})^2=f^2 e^{-2(D-d-1)b}\,.
\end{equation}

Substituting the anz\"atze and noting that $\sqrt{-g}\propto e^{da+(D-d-1)b+n}$,
we find
\begin{eqnarray}
S&\propto&
\int dt\,\, e^{da+(D-d-1)b-n}\nonumber \\
& &\quad\times\Bigl\{2 d \ddot{a}+2 (D-d-1)\ddot{b}+d(d+1)\dot{a}^2+
2d(D-d-1)\dot{a}\dot{b}+(D-d)(D-d-1)\dot{b}^2\nonumber \\
& &\qquad
-2\dot{n}\left[d\dot{a}+(D-d-1)\dot{b}\right]+(D-d-1)(D-d-2)k_be^{-2b+2n}
+\sigma\frac{1}{2}\dot{\Phi}^2\nonumber
\\ &
&\qquad-
\frac{1}{2}f^2[l e^{-2(D-d-1)b+2\kappa\alpha\Phi}
+r e^{-2(D-d-1)b-2\kappa\sigma\Phi/\alpha}]e^{2n}
\Bigr\}\,,
\end{eqnarray}
where the dot indicates the derivative with respect to time $t$.
Here, if we set
\begin{equation}
n(t)=d a(t)+(D-d-1) b(t)\,
\end{equation}
as a gauge choice, the reduced cosmological action becomes
\begin{eqnarray}
S&\propto&
\int dt \Bigl\{-d(d-1)\dot{a}^2-
2d(D-d-1)\dot{a}\dot{b}-(D-d-1)(D-d-2)\dot{b}^2+\sigma\frac{1}{2}\dot{\Phi}^2
\nonumber
\\ & &+
(D-d-1)(D-d-2)k_be^{2[da+(D-d-2)b]}-
\frac{1}{2}f^2\left[l e^{2[da+\kappa\alpha\Phi]}
+r e^{2[da-\kappa\sigma\Phi/\alpha]}\right]
\Bigr\}\,.
\label{ra}
\end{eqnarray}

We now find that the ``kinetic'' terms, which include the time derivatives, in the
reduced action (\ref{ra}) can have a unique quadratic form as follows:
\begin{eqnarray}
& &
-d(d-1)\dot{a}^2-
2d(D-d-1)\dot{a}\dot{b}-(D-d-1)(D-d-2)\dot{b}^2
+\sigma\frac{1}{2}\dot{\Phi}^2\nonumber
\\ &=&-\frac{1}{2}\frac{2(D-d-1)}{D-d-2}\left[d \dot{a}+(D-d-2)\dot{b}\right]^2
\nonumber \\
& &
+\frac{1}{2}\frac{\sigma}{\alpha^2+\sigma}\frac{2(D-2)}{d(D-d-2)}
\left[d\dot{a}+\sqrt{\frac{d(D-d-2)}{2(D-2)}}\alpha\dot{\Phi}\right]^2
\nonumber \\
& &
+\frac{1}{2}\frac{\alpha^2}{\alpha^2+\sigma}\frac{2(D-2)}{d(D-d-2)}
\left[d\dot{a}-\sqrt{\frac{d(D-d-2)}{2(D-2)}}\frac{\sigma}{\alpha}\dot{\Phi}
\right]^2
\,.
\end{eqnarray}

Therefore, if we fix the constant
\begin{equation}
\kappa\equiv\sqrt{\frac{d(D-d-2)}{2(D-2)}}\,,
\end{equation}
the action can be written in 
three independent variables $x\propto d a+(D-d-2)b$,
$y\propto d a+\kappa\alpha\Phi$
and $z\propto d a-\kappa\sigma\Phi/\alpha$.

We write the reduced action as $S=\int L\, dt$ with
\begin{eqnarray}
L&=&
-\frac{1}{2}\frac{2(D-d-1)}{D-d-2}\left[d \dot{a}+(D-d-2)\dot{b}\right]^2
+\frac{1}{2}\frac{\sigma}{\alpha^2+\sigma}\frac{2(D-2)}{d(D-d-2)}
\left[d\dot{a}+\kappa\alpha\dot{\Phi}\right]^2
\nonumber \\
& &
+\frac{1}{2}\frac{\alpha^2}{\alpha^2+\sigma}\frac{2(D-2)}{d(D-d-2)}
\left[d\dot{a}-\kappa\frac{\sigma}{\alpha}\dot{\Phi}
\right]^2\nonumber \\
& &-\frac{V_1}{2}
e^{2[da+(D-d-2)b]}-
\frac{1}{2}f^2\left[l e^{2[da+\kappa\alpha\Phi]}
+r e^{2[da-\kappa\sigma\Phi/\alpha]}\right]
\,,
\end{eqnarray}
where $V_1\equiv(D-d-1)(D-d-2)(-2k_b)$.

Note that the coefficients of $[d \dot{a}+(D-d-2)\dot{b}]^2$ in the Lagrangian $L$
are independent of the dilaton coupling $\alpha$ and the kinematical signature
$\sigma$.
Therefore, we take a variable $x$ as
\begin{equation}
x(t)\equiv\sqrt{\frac{2(D-d-1)}{D-d-2}}\left[d a+(D-d-2)b\right]\,,\\
\end{equation}
throughout this paper. Then, the variable $x(t)$ obeys the equation
\begin{equation}
\ddot{x}+\lambda_1V_1
e^{2\lambda_1x}=0\quad\mbox{with}\quad
\label{xeq}
\lambda_1\equiv
\sqrt{\frac{D-d-2}{2(D-d-1)}}\,.
\end{equation}
The solution of Eq.~(\ref{xeq}) is
\begin{itemize}
\item (negative $E_1$)
\begin{equation}
x(t)=\left\{
\begin{array}{ll}
x_{--1}(t)\equiv\frac{1}{2\lambda_1}\ln \frac{q_1^2}{\sinh^2q_1\sqrt{V_1}\lambda_1
(t-t_1)}\, & \mbox{for } k_b=-1\, \\
x_{-0}(t)\equiv q_1(t-t_1)\, & \mbox{for } k_b=0\, \\
x_{-+1}(t)\equiv\frac{1}{2\lambda_1}\ln
\frac{q_1^2}{\cosh^2q_1\sqrt{|V_1|}\lambda_1 (t-t_1)}\, & \mbox{for } k_b=+1\,, 
\end{array}
\right.
\end{equation}
where $q_1$ and $t_1$ are constants.

The first integral
$E_1\equiv-\frac{1}{2}\dot{x}^2+\frac{V_1}{2}e^{2\lambda_1x}$ is
\begin{equation}
E_1=\left\{
\begin{array}{ll}
E_{--1}\equiv-\frac{q_1^2V_1}{2}\, & \mbox{for } k_b=-1\, \\
E_{-0}\equiv-\frac{q_1^2}{2}\, & \mbox{for } k_b=0\, \\
E_{-+1}\equiv-\frac{q_1^2|V_1|}{2}\, & \mbox{for } k_b=+1\,, 
\end{array}
\right.
\end{equation}
respectively.
\item (positive $E_1$)
\begin{equation}
x(t)=x_{+-1}(t)\equiv\frac{1}{2\lambda_1}\ln
\frac{q_1^2}{\sin^2q_1\sqrt{V_1}\lambda_1 (t-t_1)}\quad\mbox{for}\quad
k_b=-1\,,
\end{equation}
where $q_1$ and $t_1$ are constants.

The first integral
$E_1=-\frac{1}{2}\dot{x}^2+\frac{V_1}{2}e^{2\lambda_1x}$ is
\begin{equation}
E_1=E_{+-1}\equiv\frac{q_1^2V_1}{2}\quad\mbox{for}\quad
k_b=-1\,.
\end{equation}
\end{itemize}

In the next section, we study the model with $\sigma=+1$ and derive exact
solutions. The case with $\sigma=-1$ is treated in Sec.~\ref{phan}.
\section{case for the canonical kinetic term of the dilaton ($\sigma=1$)}
\label{cano}
For $\sigma=+1$, the reduced cosmological Lagrangian is written as
\begin{equation}
L=-\frac{1}{2}\dot{x}^2+\frac{1}{2}\dot{y}^2+\frac{1}{2}\dot{z}^2
-\frac{V_1}{2}e^{2\lambda_1 x}-\frac{lf^2}{2}e^{2\lambda_2
y}-\frac{rf^2}{2}e^{2\lambda_3 z}\,,
\end{equation}
where
\begin{equation}
\lambda_2y\equiv d a+\kappa\alpha \Phi\,,\,
\lambda_3z\equiv d a-\frac{\kappa}{\alpha}
\Phi\quad\mbox{with}\quad
\lambda_2\equiv
\sqrt{\alpha^2+1} \,\kappa\,,\,
\lambda_3\equiv
\sqrt{{\alpha^{-2}+1}}\,\kappa\,.
\end{equation}

We permit arbitrary signs of $l$ and $r$ in the present paper.
When a negative sign of the coefficient is taken, it yields  a ``phantom'' gauge
field. Such a phantom gauge field has been considered in a cosmological
context \cite{KS}, though their negative value may lead to pathological
consequences in quantum physics.

The exact solution of the model can now be obtained in each case shown below.
In this section, we define integrals of motion as
\begin{equation}
E_2\equiv\frac{1}{2}\dot{y}^2+\frac{lf^2}{2}e^{2\lambda_2y}\,,
\quad E_3\equiv\frac{1}{2}\dot{z}^2+\frac{rf^2}{2}e^{2\lambda_3z}\,.
\end{equation}
\subsection{Case $l>0$ and $r>0$}
First, in this case,  we find that the solutions for $y$ and $z$ for the
reduced Lagrangian $L$ can be written as
\begin{equation}
y(t)=\frac{1}{2\lambda_2}\ln \frac{q_2^2}{\cosh^2q_2\sqrt{l}f\lambda_2
(t-t_2)}\,,\quad
z(t)=\frac{1}{2\lambda_3}\ln \frac{q_3^2}{\cosh^2q_3\sqrt{r}f\lambda_3
(t-t_3)}\,,
\end{equation}
where $q_2$, $t_2$, $q_3$, and $t_3$ are integration constants.
Then, both $E_2$ and
$E_3$ are positive.
Since the Hamiltonian constraint gives $E_1+E_2+E_3=0$,
a possible solution of $x(t)$ is $x_{-k_b}(t)$ defined in Sec.~\ref{av}.
The Hamiltonian constraint then reads
\begin{equation}
lf^2q_2^2+rf^2q_3^2=|V_1|q_1^2\quad\mbox{for } k_b=\pm 1\,,\quad 
lf^2q_2^2+rf^2q_3^2=q_1^2\quad\mbox{for } k_b=0\,.
\end{equation}
\subsection{Case $l>0$ and $r<0$}
In this case, the exact solution for $y$ is written as
\begin{equation}
y(t)=\frac{1}{2\lambda_2}\ln \frac{q_2^2}{\cosh^2q_2\sqrt{l}f\lambda_2
(t-t_2)}\,,
\end{equation}
where $q_2$ and $t_2$ are constants, and then $E_2>0$.
The solutions of other variables are characterized by the following subcategories,
according to the signature of $E_1$ and $E_3$.
\subsubsection{$E_1<0$ and $E_3>0$}
The solution for $x$ is $x_{-k_b}(t)$, and the solution for $z$ is
\begin{equation}
z(t)=\frac{1}{2\lambda_3}\ln \frac{q_3^2}{\sinh^2q_3\sqrt{|r|}f\lambda_3
(t-t_3)}\,,
\end{equation}
where $q_3$ and $t_3$ are constants.
Then, the Hamiltonian constraint becomes
\begin{equation}
lf^2q_2^2+|r|f^2q_3^2-|V_1|q_1^2=0\quad\mbox{for}\quad
k_b=\pm 1\,,\quad
lf^2q_2^2+|r|f^2q_3^2-q_1^2=0\quad\mbox{for}\quad
k_b=0\,.
\end{equation}
\subsubsection{$E_1<0$ and $E_3<0$}
The solution for $x$ is $x_{-k_b}(t)$, and the solution for $z$ is
\begin{equation}
z(t)=
\frac{1}{2\lambda_3}\ln \frac{q_3^2}{\sin^2q_3\sqrt{|r|}f\lambda_3
(t-t_3)}\,,
\end{equation}
where $q_3$ and $t_3$ are constants.
Then, the Hamiltonian constraint becomes
\begin{equation}
lf^2q_2^2-|r|f^2q_3^2-|V_1|q_1^2=0\quad\mbox{for}\quad
k_b=\pm 1\,,
\quad
lf^2q_2^2-|r|f^2q_3^2-q_1^2=0\quad\mbox{for}\quad
k_b=0\,.
\end{equation}
\subsubsection{$E_1>0$ and $E_3<0$}
The solution for $x$ is $x_{+-1}(t)$, and the solution for $z$ is
\begin{equation}
z(t)=
\frac{1}{2\lambda_3}\ln \frac{q_3^2}{\sin^2q_3\sqrt{|r|}f\lambda_3
(t-t_3)}\,,
\end{equation}
where $q_3$ and $t_3$ are constants.
Then, the Hamiltonian constraint becomes
\begin{equation}
lf^2q_2^2-|r|f^2q_3^2+V_1q_1^2=0\quad\mbox{for}\quad
k_b=-1\,.
\end{equation}
\subsection{Case $l<0$ and $r>0$}
This case can be regarded as the previous case where the roles of $y$ and $z$ are
mutually exchanged.
The exact solution for $z$ is written as
\begin{equation}
z(t)=\frac{1}{2\lambda_3}\ln \frac{q_3^2}{\cosh^2q_3\sqrt{r}f\lambda_3
(t-t_3)}\,,
\end{equation}
where $q_3$ and $t_3$ are constants, and then $E_3>0$.
Solutions of other variables are characterized by the following subcategories,
according to the signature of $E_1$ and $E_2$.
\subsubsection{$E_1<0$ and $E_2>0$}
The solution for $x$ is $x_{-k_b}(t)$, and the solution for $y$ is
\begin{equation}
y(t)=\frac{1}{2\lambda_2}\ln \frac{q_2^2}{\sinh^2q_2\sqrt{|l|}f\lambda_2
(t-t_2)}\,,
\end{equation}
where $q_2$ and $t_2$ are constants.
Then, the Hamiltonian constraint becomes
\begin{equation}
|l|f^2q_2^2+rf^2q_3^2-|V_1|q_1^2=0\quad\mbox{for}\quad
k_b=\pm 1\,,
\quad
|l|f^2q_2^2+rf^2q_3^2-q_1^2=0\quad\mbox{for}\quad
k_b=0\,.
\end{equation}
\subsubsection{$E_1<0$ and $E_2<0$}
The solution for $x$ is $x_{-k_b}(t)$, and the solution for $y$ is
\begin{equation}
y(t)=
\frac{1}{2\lambda_2}\ln \frac{q_2^2}{\sin^2q_2\sqrt{|l|}f\lambda_2
(t-t_2)}\,,
\end{equation}
where $q_2$ and $t_2$ are constants.
Then, the Hamiltonian constraint becomes
\begin{equation}
-|l|f^2q_2^2+rf^2q_3^2-|V_1|q_1^2=0\quad\mbox{for}\quad
k_b=\pm 1\,,
\quad
-|l|f^2q_2^2+rf^2q_3^2-q_1^2=0\quad\mbox{for}\quad
k_b=0\,.
\end{equation}
\subsubsection{$E_1>0$ and $E_2<0$}
The solution for $x$ is $x_{+-1}(t)$, and the solution for $y$ is
\begin{equation}
y(t)=
\frac{1}{2\lambda_2}\ln \frac{q_2^2}{\sin^2q_2\sqrt{|l|}f\lambda_2
(t-t_2)}\,,
\end{equation}
where $q_2$ and $t_2$ are constants.
Then, the Hamiltonian constraint becomes
\begin{equation}
-|l|f^2q_2^2+rf^2q_3^2+V_1q_1^2=0\quad\mbox{for}\quad
k_b=-1\,.
\end{equation}
\subsection{Case $l<0$ and $r<0$}
In this case, various forms of the exact solutions appear since each integral of
motion can take a positive or negative value.
\subsubsection{$E_1<0$, $E_2>0$, and $E_3>0$}
The solution for $x$ is $x_{-k_b}(t)$, and the solutions for $y$ and $z$ are
\begin{equation}
y(t)=
\frac{1}{2\lambda_2}\ln \frac{q_2^2}{\sinh^2q_2\sqrt{|l|}f\lambda_2
(t-t_2)}\,, \quad
z(t)=
\frac{1}{2\lambda_3}\ln \frac{q_3^2}{\sinh^2q_3\sqrt{|r|}f\lambda_3
(t-t_3)}\,, 
\end{equation}
where $q_2$, $t_2$, $q_3$, and $t_3$ are constants.
Then, the Hamiltonian constraint becomes
\begin{equation}
|l|f^2q_2^2+|r|f^2q_3^2-|V_1|q_1^2=0\quad\mbox{for}\quad
k_b=\pm 1\,,
\quad
|l|f^2q_2^2+|r|f^2q_3^2-q_1^2=0\quad\mbox{for}\quad
k_b=0\,.
\end{equation}
\subsubsection{$E_1<0$, $E_2>0$, and $E_3<0$}
The solution for $x$ is $x_{-k_b}(t)$, and the solutions for $y$ and $z$ are
\begin{equation}
y(t)=
\frac{1}{2\lambda_2}\ln \frac{q_2^2}{\sinh^2q_2\sqrt{|l|}f\lambda_2
(t-t_2)}\,, \quad
z(t)=
\frac{1}{2\lambda_3}\ln \frac{q_3^2}{\sin^2q_3\sqrt{|r|}f\lambda_3
(t-t_3)}\,, 
\end{equation}
where $q_2$, $t_2$, $q_3$, and $t_3$ are constants.
Then, the Hamiltonian constraint becomes
\begin{equation}
|l|f^2q_2^2-|r|f^2q_3^2-|V_1|q_1^2=0\quad\mbox{for}\quad
k_b=\pm 1\,,
\quad
|l|f^2q_2^2-|r|f^2q_3^2-q_1^2=0\quad\mbox{for}\quad
k_b=0\,.
\end{equation}
\subsubsection{$E_1<0$, $E_2<0$, and $E_3>0$}
The solution for $x$ is $x_{-k_b}(t)$, and the solutions for $y$ and $z$ are
\begin{equation}
y(t)=
\frac{1}{2\lambda_2}\ln \frac{q_2^2}{\sin^2q_2\sqrt{|l|}f\lambda_2
(t-t_2)}\,, \quad
z(t)=
\frac{1}{2\lambda_3}\ln \frac{q_3^2}{\sinh^2q_3\sqrt{|r|}f\lambda_3
(t-t_3)}\,, 
\end{equation}
where $q_2$, $t_2$, $q_3$, and $t_3$ are constants.
Then, the Hamiltonian constraint becomes
\begin{equation}
-|l|f^2q_2^2+|r|f^2q_3^2-|V_1|q_1^2=0\quad\mbox{for}\quad
k_b=\pm 1\,,
\quad
-|l|f^2q_2^2+|r|f^2q_3^2-q_1^2=0\quad\mbox{for}\quad
k_b=0\,.
\end{equation}
\subsubsection{$E_1>0$, $E_2<0$, and $E_3<0$}
The solution for $x$ is $x_{+-1}(t)$, and the solutions for $y$ and $z$ are
\begin{equation}
y(t)=
\frac{1}{2\lambda_2}\ln \frac{q_2^2}{\sin^2q_2\sqrt{|l|}f\lambda_2
(t-t_2)}\,, \quad
z(t)=
\frac{1}{2\lambda_3}\ln \frac{q_3^2}{\sin^2q_3\sqrt{|r|}f\lambda_3
(t-t_3)}\,, 
\end{equation}
where $q_2$, $t_2$, $q_3$, and $t_3$ are constants.
Then, the Hamiltonian constraint becomes
\begin{equation}
-|l|f^2q_2^2-|r|f^2q_3^2+V_1q_1^2=0\quad\mbox{for}\quad
k_b=-1\,.
\end{equation}
\subsubsection{$E_1>0$, $E_2>0$, and $E_3<0$}
The solution for $x$ is $x_{+-1}(t)$, and the solutions for $y$ and $z$ are
\begin{equation}
y(t)=
\frac{1}{2\lambda_2}\ln \frac{q_2^2}{\sinh^2q_2\sqrt{|l|}f\lambda_2
(t-t_2)}\,, \quad
z(t)=
\frac{1}{2\lambda_3}\ln \frac{q_3^2}{\sin^2q_3\sqrt{|r|}f\lambda_3
(t-t_3)}\,, 
\end{equation}
where $q_2$, $t_2$, $q_3$, and $t_3$ are constants.
Then, the Hamiltonian constraint becomes
\begin{equation}
|l|f^2q_2^2-|r|f^2q_3^2+V_1q_1^2=0\quad\mbox{for}\quad
k_b=-1\,,
\end{equation}
\subsubsection{$E_1>0$, $E_2<0$, and $E_3>0$}
The solution for $x$ is $x_{+-1}(t)$, and the solutions for $y$ and $z$ are
\begin{equation}
y(t)=
\frac{1}{2\lambda_2}\ln \frac{q_2^2}{\sin^2q_2\sqrt{|l|}f\lambda_2
(t-t_2)}\,, \quad
z(t)=
\frac{1}{2\lambda_3}\ln \frac{q_3^2}{\sinh^2q_3\sqrt{|r|}f\lambda_3
(t-t_3)}\,, 
\end{equation}
where $q_2$, $t_2$, $q_3$, and $t_3$ are constants.
Then, the Hamiltonian constraint becomes
\begin{equation}
-|l|f^2q_2^2+|r|f^2q_3^2+V_1q_1^2=0\quad\mbox{for}\quad
k_b=-1\,.
\end{equation}

\bigskip

Now, all the solutions for $x$, $y$, and $z$ have been shown for $\sigma=+1$.
In the next section, we examine the case with $\sigma=-1$,
which corresponds to a phantom dilaton.
\section{The case for a phantom dilaton ($\sigma=-1$)}
\label{phan}
In this section, we consider the case with $\sigma=-1$.
Remembering $0<\alpha<1$, the reduced action can be written as
\begin{equation}
L=-\frac{1}{2}\dot{x}^2+\frac{1}{2}\dot{y}^2-\frac{1}{2}\dot{z}^2
-\frac{V_1}{2}e^{2\lambda_1 x}-\frac{lf^2}{2}e^{2\lambda_2
y}-\frac{rf^2}{2}e^{2\lambda_3 z}\,,
\end{equation}
where
\begin{equation}
\lambda_2y\equiv d a+\kappa\alpha
\Phi\,,\, \lambda_3z\equiv d a+\frac{\kappa}{\alpha}
\Phi\quad\mbox{with}\quad
\lambda_2=
\sqrt{1-\alpha^2} \,\kappa\,,\,
\lambda_3=
\sqrt{\alpha^{-2}-1}\,\kappa\,.
\end{equation}

Similarly to the previous section,
we obtain exact solutions for the cases with positive and negative couplings $l$
and $r$.
In this section, we define
\begin{equation}
E_2\equiv\frac{1}{2}\dot{y}^2+\frac{lf^2}{2}e^{2\lambda_2y}\,,
\quad E_3\equiv-\frac{1}{2}\dot{z}^2+\frac{rf^2}{2}e^{2\lambda_3z}\,.
\end{equation}
\subsection{Case $l>0$ and $r>0$}
In this case, the solution for $y$ takes the form
\begin{equation}
y(t)=\frac{1}{2\lambda_2}\ln \frac{q_2^2}{\cosh^2q_2\sqrt{l}f\lambda_2
(t-t_2)}\,,
\end{equation}
where $q_2$ and $t_2$ are constants, and then $E_2>0$.
Similarly to the previous section, several cases are
classified below.
\subsubsection{$E_1<0$ and $E_3>0$}
The solution for $x$ is $x_{-k_b}(t)$, and the solution for $z$ is
given by
\begin{equation}
z(t)=
\frac{1}{2\lambda_3}\ln \frac{q_3^2}{\sin^2q_3\sqrt{r}f\lambda_3
(t-t_3)}\,,
\end{equation}
where $q_3$ and $t_3$ are constants.
The Hamiltonian constraint becomes
\begin{equation}
lf^2q_2^2+rf^2q_3^2-|V_1|q_1^2=0\quad\mbox{for}\quad
k_b=\pm 1\,,
\quad
lf^2q_2^2+rf^2q_3^2-q_1^2=0\quad\mbox{for}\quad
k_b=0\,.
\end{equation}
\subsubsection{$E_1<0$ and $E_3<0$}
The solution for $x$ is $x_{-k_b}(t)$, and the solutions for $z$ is
\begin{equation}
z(t)=
\frac{1}{2\lambda_3}\ln \frac{q_3^2}{\sinh^2q_3\sqrt{r}f\lambda_3
(t-t_3)}\,,
\end{equation}
where $q_3$ and $t_3$ are constants.
The Hamiltonian constraint becomes
\begin{equation}
lf^2q_2^2-rf^2q_3^2-|V_1|q_1^2=0\quad\mbox{for}\quad
k_b=\pm 1\,,
\quad
lf^2q_2^2-rf^2q_3^2-q_1^2=0\quad\mbox{for}\quad
k_b=0\,.
\end{equation}
\subsubsection{$E_1>0$ and $E_3<0$}
The solution for $x$ is $x_{+-1}(t)$, and the solution for $z$ is
\begin{equation}
z(t)=
\frac{1}{2\lambda_3}\ln \frac{q_3^2}{\sinh^2q_3\sqrt{r}f\lambda_3
(t-t_3)}\,,
\end{equation}
where $q_3$ and $t_3$ are constants.
The Hamiltonian constraint becomes
\begin{equation}
lf^2q_2^2-rf^2q_3^2+V_1q_1^2=0\quad\mbox{for}\quad
k_b=-1\,.
\end{equation}
\subsection{Case $l>0$ and $r<0$}
In this case, the solutions for $y$ and $z$ are given by
\begin{equation}
y(t)=\frac{1}{2\lambda_2}\ln \frac{q_2^2}{\cosh^2q_2\sqrt{l}f\lambda_2
(t-t_2)}\,,\quad
z(t)=
\frac{1}{2\lambda_3}\ln \frac{q_3^2}{\cosh^2q_3\sqrt{r}f\lambda_3
(t-t_3)}\,,
\end{equation}
where $q_2$, $t_2$, $q_3$, and $t_3$ are constants,
and then $E_2>$ and $E_3<0$.
\subsubsection{$E_1<0$}
The solution for $x$ is $x_{-k_b}(t)$.
The Hamiltonian constraint reads
\begin{equation}
lf^2q_2^2-|r|f^2q_3^2-|V_1|q_1^2=0\quad\mbox{for}\quad
k_b=\pm 1\,,
\quad
lf^2q_2^2-|r|f^2q_3^2-q_1^2=0\quad\mbox{for}\quad
k_b=0\,.
\end{equation}
\subsubsection{$E_1>0$}
The solution for $x$ is $x_{+-1}(t)$.
The Hamiltonian constraint reads
\begin{equation}
lf^2q_2^2-|r|f^2q_3^2+V_1q_1^2=0\quad\mbox{for}\quad
k_b=-1\,.
\end{equation}
\subsection{Case $l<0$ and $r>0$}
In this case, all possible signs for $E_1$, $E_2$, and $E_3$
can appear.
\subsubsection{$E_1<0$, $E_2>0$, and $E_3>0$}
The solution for $x$ is $x_{-k_b}(t)$.
The solutions for $y$ and $z$ are
\begin{equation}
y(t)=
\frac{1}{2\lambda_2}\ln \frac{q_2^2}{\sinh^2q_2\sqrt{|l|}f\lambda_2
(t-t_2)}\,,\quad
z(t)=
\frac{1}{2\lambda_3}\ln \frac{q_3^2}{\sin^2q_3\sqrt{r}f\lambda_3
(t-t_3)}\,.
\end{equation}
The Hamiltonian constraint is
\begin{equation}
|l|f^2q_2^2+rf^2q_3^2-|V_1|q_1^2=0\quad\mbox{for}\quad
k_b=\pm 1\,,
\quad
|l|f^2q_2^2+rf^2q_3^2-q_1^2=0\quad\mbox{for}\quad
k_b=0\,.
\end{equation}
\subsubsection{$E_1<0$, $E_2>0$, and $E_3<0$}
The solution for $x$ is $x_{-k_b}(t)$.
The solutions for $y$ and $z$ are
\begin{equation}
y(t)=
\frac{1}{2\lambda_2}\ln \frac{q_2^2}{\sinh^2q_2\sqrt{|l|}f\lambda_2
(t-t_2)}\,,\quad
z(t)=
\frac{1}{2\lambda_3}\ln \frac{q_3^2}{\sinh^2q_3\sqrt{r}f\lambda_3
(t-t_3)}\,.
\end{equation}
The Hamiltonian constraint is
\begin{equation}
|l|f^2q_2^2-rf^2q_3^2-|V_1|q_1^2=0\quad\mbox{for}\quad
k_b=\pm 1\,,
\quad
|l|f^2q_2^2-rf^2q_3^2-q_1^2=0\quad\mbox{for}\quad
k_b=0\,.
\end{equation}
\subsubsection{$E_1<0$, $E_2<0$, and $E_3>0$}
The solution for $x$ is $x_{-k_b}(t)$.
The solutions for $y$ and $z$ are
\begin{equation}
y(t)=
\frac{1}{2\lambda_2}\ln \frac{q_2^2}{\sin^2q_2\sqrt{|l|}f\lambda_2
(t-t_2)}\,,\quad
z(t)=
\frac{1}{2\lambda_3}\ln \frac{q_3^2}{\sin^2q_3\sqrt{r}f\lambda_3
(t-t_3)}\,.
\end{equation}
The Hamiltonian constraint is
\begin{equation}
-|l|f^2q_2^2+rf^2q_3^2-|V_1|q_1^2=0\quad\mbox{for}\quad
k_b=\pm 1\,,
\quad
-|l|f^2q_2^2+rf^2q_3^2-q_1^2=0\quad\mbox{for}\quad
k_b=0\,.
\end{equation}
\subsubsection{$E_1>0$, $E_2>0$, and $E_3<0$}
The solution for $x$ is $x_{+-1}(t)$.
The solutions for $y$ and $z$ are
\begin{equation}
y(t)=
\frac{1}{2\lambda_2}\ln \frac{q_2^2}{\sinh^2q_2\sqrt{|l|}f\lambda_2
(t-t_2)}\,,\quad
z(t)=
\frac{1}{2\lambda_3}\ln \frac{q_3^2}{\sinh^2q_3\sqrt{r}f\lambda_3
(t-t_3)}\,.
\end{equation}
The Hamiltonian constraint is
\begin{equation}
|l|f^2q_2^2-rf^2q_3^2+V_1q_1^2=0\quad\mbox{for}\quad
k_b=-1\,.
\end{equation}
\subsubsection{$E_1>0$, $E_2<0$, and $E_3>0$}
The solution for $x$ is $x_{+-1}(t)$.
The solutions for $y$ and $z$ are
\begin{equation}
y(t)=
\frac{1}{2\lambda_2}\ln \frac{q_2^2}{\sin^2q_2\sqrt{|l|}f\lambda_2
(t-t_2)}\,,\quad
z(t)=
\frac{1}{2\lambda_3}\ln \frac{q_3^2}{\sin^2q_3\sqrt{r}f\lambda_3
(t-t_3)}\,.
\end{equation}
The Hamiltonian constraint is
\begin{equation}
-|l|f^2q_2^2+rf^2q_3^2+V_1q_1^2=0\quad\mbox{for}\quad
k_b=-1\,.
\end{equation}
\subsubsection{$E_1>0$, $E_2<0$, and $E_3<0$}
The solution for $x$ is $x_{+-1}(t)$.
The solutions for $y$ and $z$ are
\begin{equation}
y(t)=
\frac{1}{2\lambda_2}\ln \frac{q_2^2}{\sin^2q_2\sqrt{|l|}f\lambda_2
(t-t_2)}\,,\quad
z(t)=
\frac{1}{2\lambda_3}\ln \frac{q_3^2}{\sinh^2q_3\sqrt{r}f\lambda_3
(t-t_3)}\,.
\end{equation}
The Hamiltonian constraint is
\begin{equation}
-|l|f^2q_2^2-rf^2q_3^2+V_1q_1^2=0\quad\mbox{for}\quad
k_b=-1\,.
\end{equation}
\subsection{Case $l<0$ and $r<0$}
In this case, the solution for $z$ is
\begin{equation}
z(t)=
\frac{1}{2\lambda_3}\ln \frac{q_3^2}{\cosh^2q_3\sqrt{|r|}f\lambda_3
(t-t_3)}\,,
\end{equation}
and then $E_3<0$.
\subsubsection{$E_1<0$ and $E_2>0$}
The solution for $x$ is $x_{-k_b}(t)$, and the solution for $y$ is
\begin{equation}
y(t)=
\frac{1}{2\lambda_2}\ln \frac{q_2^2}{\sinh^2q_2\sqrt{|l|}f\lambda_2
(t-t_2)}\,,
\end{equation}
where $q_2$ and $t_2$ are constants.
The Hamiltonian constraint reads
\begin{equation}
|l|f^2q_2^2-|r|f^2q_3^2-|V_1|q_1^2=0\quad\mbox{for}\quad
k_b=\pm 1\,,
\quad
|l|f^2q_2^2-|r|f^2q_3^2-q_1^2=0\quad\mbox{for}\quad
k_b=0\,.
\end{equation}
\subsubsection{$E_1>0$ and $E_2>0$}
The solution for $x$ is $x_{+-1}(t)$, and the solution for $y$ is
\begin{equation}
y(t)=
\frac{1}{2\lambda_2}\ln \frac{q_2^2}{\sinh^2q_2\sqrt{|l|}f\lambda_2
(t-t_2)}\,,
\end{equation}
where $q_2$ and $t_2$ are constants.
The Hamiltonian constraint reads
\begin{equation}
|l|f^2q_2^2-|r|f^2q_3^2+V_1q_1^2=0\quad\mbox{for}\quad
k_b=-1\,.
\end{equation}
\subsubsection{$E_1>0$ and $E_2<0$}
The solution for $x$ is $x_{+-1}(t)$, and the solution for $y$ is
\begin{equation}
y(t)=
\frac{1}{2\lambda_2}\ln \frac{q_2^2}{\sin^2q_2\sqrt{|l|}f\lambda_2
(t-t_2)}\,,
\end{equation}
where $q_2$ and $t_2$ are constants.
The Hamiltonian constraint reads
\begin{equation}
-|l|f^2q_2^2-|r|f^2q_3^2+V_1q_1^2=0\quad\mbox{for}\quad
k_b=-1\,.
\end{equation}
\section{Accelerating universe}
\label{ac}
To analyze the cosmological behavior closely, we introduce the
``physical '' $(d+1)$-dimensional metric and the cosmic time.
When we take a representation for the $D$-dimensional metric such as
\begin{eqnarray}
ds^2&=&e^{-\frac{2(D-d-1)b}{d-1}}\bar{g}_{\mu\nu}dx^\mu dx^\nu+
e^{2b}\tilde{g}_{mn}dx^mdx^n\,,
\end{eqnarray}
we find that
$\sqrt{-g}R$ is proportional to $\sqrt{-\bar{g}}\bar{R}+\cdots$, where $\bar{R}$ is
the Ricci scalar of the $(d+1)$-dimensional spacetime constructed from
$\bar{g}_{\mu\nu}$.
Therefore, the metric $\bar{g}_{\mu\nu}$
is considered to define the Einstein frame of the $(d+1)$-dimensional spacetime.

In the present study, we should regard the following form for the metric:
\begin{eqnarray}
ds^2&=&-e^{2[da(t)+(D-d-1)b(t)]}dt^2+e^{2a(t)}d\mathbf{x}^2+
e^{2b(t)}d\Omega_{D-d-1}^2\nonumber\\
&=&e^{-\frac{2(D-d-1)b}{d-1}}(-d\eta^2+S^2(\eta)d\mathbf{x}^2)+
e^{2b}d\Omega_{D-d-1}^2\,,
\end{eqnarray}
where $\eta$ is the cosmic time for the $(d+1)$-dimensional spacetime
and $S$ is the ``physical'' scale factor of $d$-dimensional flat space in
the $(d+1)$-dimensional view. Thus, we obtain the relations
\begin{equation}
S(\eta)=e^{a(t)+\frac{D-d-1}{d-1}b(t)}\,,\quad
d\eta=\pm e^{d[a(t)+\frac{D-d-1}{d-1}b(t)]}dt=\pm S^d dt\,.
\end{equation}
They can be written in terms of $x$, $y$, and $z$ as follows:
\begin{eqnarray}
& &S=\left(e^{2\lambda_1x}\right)^{\frac{D-d-1}{2(d-1)(D-d-2)}}
\left(e^{2\lambda_2y}\right)^{-\frac{D-2}{2(\alpha^2+1)d(d-1)(D-d-2)}}
\left(e^{2\lambda_3z}\right)^{-\frac{\alpha^2(D-2)}{2(\alpha^2+1)d(d-1)(D-d-2)}}
\,\mbox{for}\,\sigma=1\,,\\
& &S=\left(e^{2\lambda_1x}\right)^{\frac{D-d-1}{2(d-1)(D-d-2)}}
\left(e^{2\lambda_2y}\right)^{-\frac{D-2}{2(1-\alpha^2)d(d-1)(D-d-2)}}
\left(e^{2\lambda_3z}\right)^{\frac{\alpha^2(D-2)}{2(1-\alpha^2)d(d-1)(D-d-2)}}
\,\mbox{for}\,\sigma=-1\,.
\end{eqnarray}
\subsection{Some special solutions expressed by elementary functions of $\eta$}
Unfortunately, the solutions listed in the previous sections and
Appendix~\ref{ss} cannot be written in terms of elementary functions of $\eta$ in
general. There are, however, special cases where the solutions can be expressed
as simple functions of $\eta$. We first consider these cases.
\begin{itemize}
\item $\sigma=1$, $l<0$, $r<0$, and $k_b=-1$
(\ref{+---;-++}, \ref{+---;-+-}, \ref{+---;--+},
\ref{+---;+--}, \ref{+---;++-}, \ref{+---;+-+})

Taking the limit of $q\rightarrow 0$, we find
\begin{eqnarray}
& &e^{2da(t)}=\frac{1}{f^2|l\lambda_2^2|^{1/(1+\alpha^2)}
|r\lambda_3^2|^{\alpha^2/(1+\alpha^2)}}%
\left(t-t_2\right)^{-\frac{2}{\alpha^2+1}}
\left(t-t_3\right)^{-\frac{2\alpha^2}{\alpha^2+1}}\,,
\\& &e^{2(D-d-2)b(t)}=\frac{f^2|l\lambda_2^2|^{1/(1+\alpha^2)}
|r\lambda_3^2|^{\alpha^2/(1+\alpha^2)}}%
{V_1\lambda_1^2(t-t_1)^{2}}
\left(t-t_2\right)^{\frac{2}{\alpha^2+1}}
\left(t-t_3\right)^{\frac{2\alpha^2}{\alpha^2+1}}\,,
\\& &e^{2\kappa\Phi(t)}=\left(\sqrt{\frac{|r|}{\alpha^2|l|}}
\frac{t-t_3}{t-t_2}
\right)^{\frac{2\alpha}{\alpha^2+1}}
\,.
\end{eqnarray}
Further, if $t_1=t_2=t_3\equiv t_0$, they become
\begin{equation}
e^{2da(t)}=\frac{1}{C^2(t-t_0)^2}\,,~~
e^{2(D-d-2)b}=\frac{C^2}%
{V_1\lambda_1^2}\,,~~
e^{2\kappa\Phi}=\left(\frac{|r|}{\alpha^2|l|}
\right)^{\frac{\alpha}{\alpha^2+1}}
\,.
\end{equation}
with $C^2\equiv f^2|l\lambda_2^2|^{1/(1+\alpha^2)}|r\lambda_3^2|
^{\alpha^2/(1+\alpha^2)}$.
Then we obtain
\begin{equation}
S(\eta)\propto\exp\left[\frac{1}{d}C^{-\frac{D-2}{(d-1)(D-d-2)}}\eta\right]\,.
\end{equation}
This solution describes $(d+1)$-dimensional de Sitter spacetime.

\item $\sigma=1$, $l<0$, $r<0$, and $k_b=0$ (\ref{+--0;-++}, \ref{+--0;-+-},
\ref{+--0;--+})

Taking the limit of $q\rightarrow 0$, we find
\begin{eqnarray}
& &e^{2da(t)}=\frac{1}{f^2|l\lambda_2^2|^{1/(1+\alpha^2)}
|r\lambda_3^2|^{\alpha^2/(1+\alpha^2)}}%
\left(t-t_2\right)^{-\frac{2}{\alpha^2+1}}
\left(t-t_3\right)^{-\frac{2\alpha^2}{\alpha^2+1}}\,,
\\& &e^{2(D-d-2)b(t)}=C_1{f^2|l\lambda_2^2|^{1/(1+\alpha^2)}
|r\lambda_3^2|^{\alpha^2/(1+\alpha^2)}}%
\left(t-t_2\right)^{\frac{2}{\alpha^2+1}}
\left(t-t_3\right)^{\frac{2\alpha^2}{\alpha^2+1}}\,,
\\& &e^{2\kappa\Phi(t)}=\left(\sqrt{\frac{|r|}{\alpha^2|l|}}
\frac{t-t_3}{t-t_2}
\right)^{\frac{2\alpha}{\alpha^2+1}}
\,,
\end{eqnarray}
where $C_1$ is a constant.
Further, if $t_2=t_3\equiv t_0$, they become
\begin{equation}
e^{2da(t)}=\frac{1}{C^2(t-t_0)^2}\,,~~
e^{2(D-d-2)b}=C_1C^2(t-t_0)^2\,,~~
e^{2\kappa\Phi}=\left(\frac{|r|}{\alpha^2|l|}
\right)^{\frac{\alpha}{\alpha^2+1}}
\,,
\end{equation}
with $C^2\equiv f^2|l\lambda_2|^{1/(1+\alpha^2)}|r\lambda_3|
^{\alpha^2/(1+\alpha^2)}$.
Then we obtain
\begin{equation}
S(\eta)\propto (\eta-\eta_0)^{\frac{D-2}{d^2(D-d-1)}}\,,\quad
e^{b}\propto (\eta-\eta_0)^{\frac{d-1}{d(D-d-1)}}\,,
\end{equation}
where $\eta_0$ is a constant.
\item $\sigma=-1$, $l<0$, $r>0$, and $k_b=-1$ (\ref{--+-;-++},
\ref{--+-;-+-}, \ref{--+-;--+}, \ref{--+-;++-}, \ref{--+-;+-+},
\ref{--+-;+--})

Taking the limit of $q\rightarrow 0$, we find
\begin{eqnarray}
& &e^{2da(t)}=\frac{1}{f^2|l\lambda_2^2|^{1/(1-\alpha^2)}
|r\lambda_3^2|^{-\alpha^2/(1-\alpha^2)}}%
\left(t-t_2\right)^{-\frac{2}{1-\alpha^2}}
\left(t-t_3\right)^{\frac{2\alpha^2}{1-\alpha^2}}\,,
\\& &e^{2(D-d-2)b(t)}=\frac{f^2|l\lambda_2^2|^{1/(1-\alpha^2)}
(r\lambda_3^2)^{-\alpha^2/(1-\alpha^2)}}%
{V_1\lambda_1^2(t-t_1)^{2}}
\left(t-t_2\right)^{\frac{2}{1-\alpha^2}}
\left(t-t_3\right)^{-\frac{2\alpha^2}{1-\alpha^2}}\,,
\\& &e^{2\kappa\Phi(t)}=\left(\sqrt{\frac{r}{\alpha^2|l|}}
\frac{t-t_3}{t-t_2}
\right)^{-\frac{2\alpha}{1-\alpha^2}}
\,.
\end{eqnarray}
Further, if $t_1=t_2=t_3\equiv t_0$, they become
\begin{equation}
e^{2da(t)}=\frac{1}{C'^2(t-t_0)^2}\,,~~
e^{2(D-d-2)b}=\frac{C'^2}%
{V_1\lambda_1^2}\,,~~
e^{2\kappa\Phi}=\left(\frac{r}{\alpha^2|l|}
\right)^{-\frac{\alpha}{1-\alpha^2}}
\,.
\end{equation}
with $C'^2\equiv f^2|l\lambda_2^2|^{1/(1-\alpha^2)}(r\lambda_3^2)
^{-\alpha^2/(1-\alpha^2)}$.
Then we obtain
\begin{equation}
S(\eta)\propto\exp\left[\frac{1}{d}C'^{-\frac{D-2}{(d-1)(D-d-2)}}\eta\right]\,.
\end{equation}
This solution describes $(d+1)$-dimensional de Sitter spacetime.
\item $\sigma=-1$, $l<0$, $r>0$, and $k_b=0$ (\ref{--+0;-++},
\ref{--+0;-+-}, \ref{--+0;--+})

Taking the limit of $q\rightarrow 0$, we find
\begin{eqnarray}
& &e^{2da(t)}=\frac{1}{f^2|l\lambda_2^2|^{1/(1-\alpha^2)}
(r\lambda_3^2)^{-\alpha^2/(1-\alpha^2)}}%
\left(t-t_2\right)^{-\frac{2}{1-\alpha^2}}
\left(t-t_3\right)^{\frac{2\alpha^2}{1-\alpha^2}}\,,
\\& &e^{2(D-d-2)b(t)}=C_1{f^2|l\lambda_2^2|^{1/(1-\alpha^2)}
(r\lambda_3^2)^{-\alpha^2/(1-\alpha^2)}}%
\left(t-t_2\right)^{\frac{2}{1-\alpha^2}}
\left(t-t_3\right)^{-\frac{2\alpha^2}{1-\alpha^2}}\,,
\\& &e^{2\kappa\Phi(t)}=\left(\sqrt{\frac{r}{\alpha^2|l|}}
\frac{t-t_3}{t-t_2}
\right)^{-\frac{2\alpha}{1-\alpha^2}}
\,.
\end{eqnarray}
Further, if $t_2=t_3\equiv t_0$, they become
\begin{equation}
e^{2da(t)}=\frac{1}{C'^2(t-t_0)^2}\,,~~
e^{2(D-d-2)b}=C_1C'^2(t-t_0)^2\,,~~
e^{2\kappa\Phi}=\left(\frac{r}{\alpha^2|l|}
\right)^{-\frac{\alpha}{1-\alpha^2}}
\,,
\end{equation}
with $C'^2\equiv f^2|l\lambda_2|^{1/(1-\alpha^2)}(r\lambda_3)
^{-\alpha^2/(1-\alpha^2)}$.
Then, we obtain
\begin{equation}
S(\eta)\propto (\eta-\eta_0)^{\frac{D-2}{d^2(D-d-1)}}\,,\quad
e^{b}\propto (\eta-\eta_0)^{\frac{d-1}{d(D-d-1)}}\,,
\end{equation}
where $\eta_0$ is a constant.
\end{itemize}
In all the cases examined above, the dilaton field $\Phi$ takes a constant value in
the final consideration, which corresponds to the extremum of the effective scalar
potential coming from the coupling to the constant fluxes, which is proportional
to $le^{2\kappa\alpha\Phi}+re^{-2\kappa\sigma\Phi/\alpha}$.

The cases with ($\sigma=1$, $l<0$, $r<0$, $k_b=-1$) and 
($\sigma=-1$, $l<0$, $r>0$, $k_b=-1$) yield the exponentially expanding universe.
Unfortunately, these are not the most general cosmological solutions of the model.
In the next subsection, we consider asymptotic behaviors of the physical scale
factor $S(\eta)$.
\subsection{Asymptotic behaviors of solutions}
We have several cases where the asymptotic behavior of $S(\eta)$, as a function of
the cosmic time $\eta$, can be obtained.
If $S(t)\sim e^{Bt}$ with constant $B$, $S(\eta)\sim \eta^{1/d}$. 
If $S(t)\sim |t-t_0|^\beta$, $S(\eta)\sim|\eta-\eta_0|^{\beta/(d\beta+1)}$
with some constants $t_0$ and $\eta_0$. The scale of the compact space $e^b$
generally shows a similar behavior to $S(\eta)$.

Thus, some typical cases, in which the solution represents the expanding
universe, can be found as follows:
\begin{itemize}
\item
$S(\eta)\sim \eta^{1/d}\quad\mbox{for}\quad \eta\rightarrow 0\,,\quad
S(\eta)\sim \eta^{\frac{D-d-1}{D-2}}\quad\mbox{for}\quad \eta\rightarrow
\infty\,.$

This behavior is found in the cases
\begin{tabbing}
$\sigma=1$, $l>0$, $r>0$, $k_b=-1$\,,\\
$\sigma=1$, $l>0$, $r<0$, $k_b=+1$, ($E_1<0$, $E_2>0$, $E_3>0$), \\ 
$\sigma=1$, $l>0$, $r<0$, $k_b=-1$, ($E_1<0$, $E_2>0$, $E_3>0$), $t_1<t_3$\,,\\
$\sigma=1$, $l<0$, $r>0$, $k_b=-1$, ($E_1<0$, $E_2>0$, $E_3>0$), $t_1<t_2$\,,\\
$\sigma=1$, $l<0$, $r<0$, $k_b=-1$, ($E_1<0$, $E_2>0$, $E_3>0$), 
$t_1<\mbox{Min}(t_2,t_3)$\,.
\end{tabbing}
\item $S(\eta)\sim \eta^{1/d}.$

This behavior is found in the cases
\begin{tabbing}
$\sigma=1$, $l>0$, $r>0$, $k_b=0, +1$\,,\\ 
$\sigma=1$, $l>0$, $r<0$, $k_b=0$ ($E_1<0$, $E_2>0$, $E_3>0$)\,,\\
$\sigma=1$, $l<0$, $r>0$, $k_b=0$ ($E_1<0$, $E_2>0$, $E_3>0$)\,.
\end{tabbing}

\item
$S(\eta)\sim
\eta^{\frac{\alpha^2(D-2)}{\alpha^2d(D-2)+(\alpha^2+1)(d-1)(D-d-2)}}\quad\mbox{for}\quad
\eta\rightarrow 0\,,\quad S(\eta)\sim
\eta^{\frac{D-d-1}{D-2}}\quad\mbox{for}\quad
\eta\rightarrow
\infty\,.$

This behavior is found in the cases
\begin{tabbing}
$\sigma=1$, $l>0$, $r<0$, $k_b=-1$, ($E_1<0$, $E_2>0$, $E_3<0$), $t_1<t_3$\,,\\
$\sigma=1$, $l<0$, $r<0$, $k_b=-1$, ($E_1<0$, $E_2>0$, $E_3<0$),
$t_1<\mbox{Min}(t_2,t_3)$\,.
\end{tabbing}

\item
$S(\eta)\sim
\eta^{\frac{\alpha^2(D-2)}{\alpha^2d(D-2)+(\alpha^2+1)(d-1)(D-d-2)}}$.

This behavior is found in the cases
\begin{tabbing}
$\sigma=1$, $l>0$, $r<0$, $k_b=0, +1$, ($E_1<0$, $E_2>0$, $E_3<0$),\\
$\sigma=1$, $l>0$, $r<0$, $k_b=-1$, ($E_1<0$, $E_2>0$, $E_3<0$), $t_1>t_3$\,,\\
$\sigma=1$, $l<0$, $r<0$, $k_b=0, \pm 1$, ($E_1<0$, $E_2>0$, $E_3<0$),
$t_1>t_2$, $t_3>t_2$\,.
\end{tabbing}

\item
$S(\eta)\sim\eta^{1/d}\quad\mbox{for}\quad
\eta\rightarrow 0\,,\quad S(\eta)\sim
\eta^{\frac{\alpha^2(D-2)}{\alpha^2d(D-2)+(\alpha^2+1)(d-1)(D-d-2)}}\quad\mbox{for}\quad
\eta\rightarrow
\infty$.

This behavior is found in the case
$\sigma=1$, $l>0$, $r<0$, $k_b=-1$, ($E_1<0$, $E_2>0$, $E_3<0$),
$t_1>t_3$\,.

\item
$S(\eta)\sim
\eta^{\frac{(D-2)}{d(D-2)+(\alpha^2+1)(d-1)(D-d-2)}}\quad\mbox{for}\quad
\eta\rightarrow 0\,,\quad S(\eta)\sim
\eta^{\frac{D-d-1}{D-2}}\quad\mbox{for}\quad
\eta\rightarrow
\infty\,.$

This behavior is found in the cases
\begin{tabbing}
$\sigma=1$, $l<0$, $r>0$, $k_b=-1$, ($E_1<0$, $E_2<0$, $E_3>0$), $t_1<t_2$\,,\\
$\sigma=1$, $l<0$, $r<0$, $k_b=-1$, ($E_1<0$, $E_2>0$, $E_3<0$),
$t_1<\mbox{Min}(t_2,t_3)$\,.
\end{tabbing}

\item
$S(\eta)\sim
\eta^{\frac{(D-2)}{d(D-2)+(\alpha^2+1)(d-1)(D-d-2)}}$.

This behavior is found in the cases
\begin{tabbing}
$\sigma=1$, $l<0$, $r>0$, $k_b=0, +1$, ($E_1<0$, $E_2<0$, $E_3>0$),\\
$\sigma=1$, $l<0$, $r>0$, $k_b=-1$, ($E_1<0$, $E_2<0$, $E_3>0$),
$t_1>t_2$\,.
\end{tabbing}

\item
$S(\eta)\sim\eta^{1/d}\quad\mbox{for}\quad\eta\rightarrow 0,\quad S(\eta)\sim
\eta^{\frac{(D-2)}{d(D-2)+(\alpha^2+1)(d-1)(D-d-2)}}\quad\mbox{for}\quad\eta
\rightarrow\infty$.

This behavior is found in the case
$\sigma=1$, $l<0$, $r>0$, $k_b=-1$, ($E_1<0$, $E_2>0$, $E_3>0$),
$t_1>t_2$\,.

\item
$S(\eta)\sim\eta^{\frac{\alpha^2(D-2)}{\alpha^2d(D-2)+(\alpha^2+1)(d-1)(D-d-2)}}\quad\mbox{for}\quad\eta\rightarrow
0,\quad S(\eta)\sim
\eta^{\frac{(D-2)}{d(D-2)+(\alpha^2+1)(d-1)(D-d-2)}}\quad\mbox{for}\quad\eta
\rightarrow\infty$.

This behavior is found in the case
$\sigma=1$, $l<0$, $r<0$, $k_b=0,\pm 1$, ($E_1<0$, $E_2>0$, $E_3<0$),
$t_1>t_3$, $t_2>t_3$\,.

\item
$S(\eta)\sim\eta^{\frac{\alpha^2(D-2)}{\alpha^2d(D-2)+(\alpha^2+1)(d-1)(D-d-2)}}\quad\mbox{for}\quad\eta\rightarrow
0,\quad S(\eta)\sim
\eta^{\frac{(D-2)}{d^2(D-d-1)}}\quad\mbox{for}\quad\eta
\rightarrow\infty$.

This behavior is found in the case
$\sigma=1$, $l<0$, $r<0$, $k_b=0,\pm 1$, ($E_1<0$, $E_2>0$, $E_3<0$),
$t_1>t_2=t_3$\,.

\item
$S(\eta)\sim \eta^{1/d}\quad\mbox{for}\quad \eta\rightarrow 0\,,\quad
S(\eta)\sim
\eta^{\frac{(d-1)(D-d-2)}{d(D-d-1)}\alpha^2+\frac{D-d-1}{D-2}}\quad\mbox{for}\quad
\eta\rightarrow
\infty\,.$

This behavior is found in the case
$\sigma=1$, $l>0$, $r<0$, $k_b=-1$, $t_1=t_3$.

\item
$S(\eta)\sim \eta^{1/d}\quad\mbox{for}\quad \eta\rightarrow 0\,,\quad
S(\eta)\sim
\eta^{\frac{(d-1)(D-d-2)}{d(D-d-1)}\alpha^{-2}+\frac{D-d-1}{D-2}}\quad\mbox{for}\quad
\eta\rightarrow
\infty\,.$

This behavior is found in the case
$\sigma=1$, $l<0$, $r>0$, $k_b=-1$, ($E_1<0$, $E_2>0$, $E_3>0$), $t_1=t_2$.

\item Expanding and contracting in a finite cosmic time $\eta$.
\begin{tabbing}
$\sigma=1$, $l>0$, $r>0$, $k_b=+1$\,,\\
$\sigma=1$, $l>0$, $r<0$, $k_b=-1$, ($E_1<0$, $E_2>0$, $E_3>0$), $t_1>t_3$\,,\\
$\sigma=1$, $l>0$, $r<0$, $k_b=-1$, ($E_1<0$, $E_2>0$, $E_3<0$), $t_1>t_3$\,,\\
$\sigma=1$, $l>0$, $r<0$, $k_b=+1$ ($E_1<0$, $E_2>0$, $E_3>0$)\,,\\
$\sigma=1$, $l>0$, $r<0$, $k_b=0,+1$ ($E_1<0$, $E_2>0$, $E_3<0$)\\
$\sigma=1$, $l<0$, $r>0$, $k_b=-1$, ($E_1<0$, $E_2>0$, $E_3>0$), $t_1>t_2$\,,\\
$\sigma=1$, $l<0$, $r>0$, $k_b=-1$, ($E_1<0$, $E_2>0$, $E_3<0$), $t_1>t_2$\,,\\
$\sigma=1$, $l<0$, $r>0$, $k_b=+1$ ($E_1<0$, $E_2>0$, $E_3>0$)\,,\\
$\sigma=1$, $l<0$, $r>0$, $k_b=0,+1$ ($E_1<0$, $E_2<0$, $E_3>0$)\,,\\
$\sigma=1$, $l<0$, $r<0$, $k_b=-1,0,+1$, ($E_1<0$),
$t_1>\mbox{Max}(t_2,t_3)$\,.
\end{tabbing}
\end{itemize}

In our model, the eternally expanding universe can be found, $k_b=-1$ ($\sigma=1$),
similar to the cases with the models studied in \cite{TW,Ohta2,Roy}.

Here we omitted analyses on interesting cases with $E_1>0$ and/or $\sigma=-1$,
in which complicated evolutions, including a bouncing universe, can be found
accordingly, to a proper choice of integration constants.
We leave the broad study of such cases for future work, and 
we restrict ourselves to considering a possible accelerating phase in the universe
in the rest of this section.

\subsection{Transient acceleration and the scalar field value moving in a finite
range}

We cannot determine the existence or absence of transient acceleration only from
asymptotic behavior. Therefore, we should investigate the behavior of $S(\eta)$
in the intermediate era
more closely. To this end, we first observe
\begin{equation}
\frac{dS}{d\eta}=S^{-d}\frac{dS}{dt}=
-\frac{1}{d-1}\frac{dS^{1-d}}{dt}\,,\quad
\frac{d^2S}{d\eta^2}=-\frac{1}{d-1}S^{-d}\frac{d^2S^{1-d}}{dt^2}\,.
\end{equation}
Thus, for expanding and accelerating physical universes,
$-\frac{dS^{1-d}}{dt}>0$ and $-\frac{d^2S^{1-d}}{dt^2}>0$.

As already known, the model with $k_b=-1$, i.e., with the hyperbolic internal
space, yields an accelerating universe in both cases with no other field
content \cite{TW,Ohta2} and with the single flux and the dilaton field \cite{Roy}.
In our model, therefore, a transient acceleration occurs for a wide range of
parameters.

The minute behavior of $S(\eta)$ is diverse in many solutions.
We concentrate mainly on the case $\sigma=1$, $l>0$, $r>0$, $k_b=-1,0$
here, not only because this case yields an expanding universe but also because this
is the only admissible case for quantum field theory, in a naive sense.
We will, however, add a discussion on a special phantom case in the last
subsection.

One of the most remarkable features of this case in our model is that the value of
the dilaton scalar field can be finite throughout the evolution of the universe
because of two dilaton couplings to fluxes.
Indeed, for the case ($\sigma=1$, $l>0$, and $r>0$), we find
\begin{equation}
e^{2\kappa\Phi(t)}=\left(\frac{r\cos^2\theta}{l\sin^2\theta}
\frac{\cosh^2[q\sin\theta
\lambda_3(t-t_3)]}{\cosh^2[q\cos\theta
\lambda_2(t-t_2)]}
\right)^{\frac{\alpha}{\alpha^2+1}}\,,
\end{equation}
where $q$, $\theta$ , $t_2$, and $t_3$ are integration constants.
When $\lambda_2\cos\theta=\lambda_3\sin\theta$, i.e.,
$\cos\theta=\lambda_3/\sqrt{\lambda_2^2+\lambda_3^2}=1/\sqrt{1+\alpha^2}$ and
$\sin\theta=\lambda_2/\sqrt{\lambda_2^2+\lambda_3^2}=\alpha/\sqrt{1+\alpha^2}$,
the scalar field $\Phi$ behaves as
\begin{equation}
e^{2\kappa\left(\alpha+\frac{1}{\alpha}\right)\Phi(t)}=\frac{r}{l\alpha^2}
\frac{\cosh^2[q\kappa(t-t_3)]}{\cosh^2[q\kappa(t-t_2)]}\,.
\end{equation}
Obviously, this is only the case of finite $\Phi(t)$ at $t\rightarrow\mp\infty$,
whose value is given by
$\lim_{t\rightarrow\mp\infty}e^{2\kappa\left(\alpha+\frac{1}{\alpha}\right)\Phi(t)}
=\frac{r}{l\alpha^2}e^{\mp 2q\kappa(t_2-t_3)}$. Note that for the case with
$k_b=-1$, since $\eta\rightarrow\infty$ at $t=t_1$, the value of $\Phi$ approaches
the constant $\Phi(t_1)$, while other variables move as $S(\eta)\sim
\eta^{\frac{D-d-1}{D-2}}$ and
$e^b\sim
\eta^{\frac{d-1}{D-2}}$. In particular, if we choose
$t_2=t_3$,
$\Phi$ takes a constant value, which is the equilibrium point of the reduced
potential
$\propto le^{2\kappa\alpha\Phi}+re^{-2\kappa\Phi/\alpha}$ included in
the action.

\begin{figure}[ht]
\centering
\includegraphics[width=5cm]{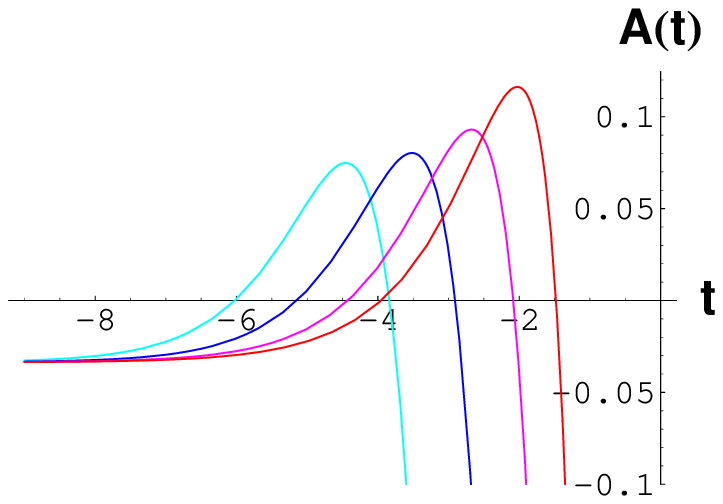}
\includegraphics[width=5cm]{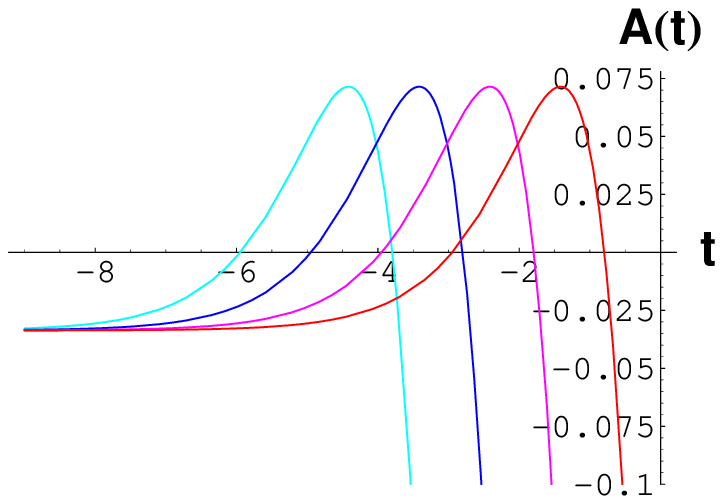}
\includegraphics[width=5cm]{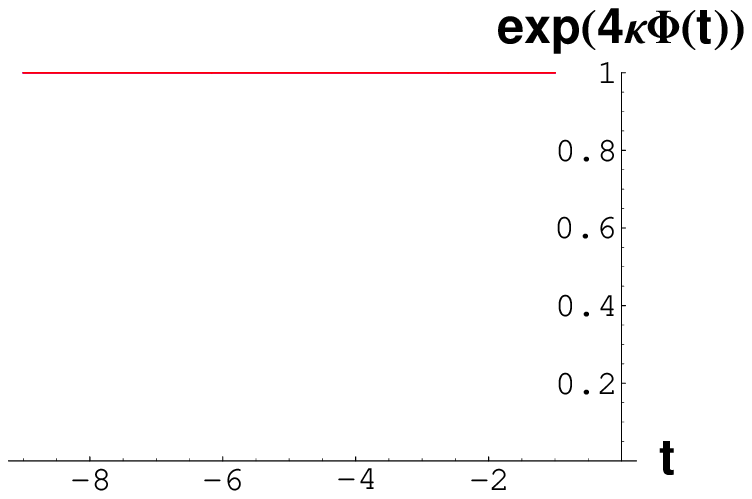}
\\
(a) \hspace{4cm} (b) \hspace{4cm} (c)
\caption{(a) $A(t)$ for $k_b=-1$ as a function of $t$ in the canonical case. 
The curves correspond to the
cases with
$t_2=t_3=-3, -2, -1, 0$, according to the location of the peak from left to right.
(b) $A(t)$ for $k_b=0$ as a function of $t$. The choice of parameters is the same
as (a). (c) $\exp(4\kappa\Phi(t))=1$ is constant in this case.
For the other parameters, see text.}
\label{fig1}
\end{figure}

\begin{figure}[ht]
\centering
\includegraphics[width=5cm]{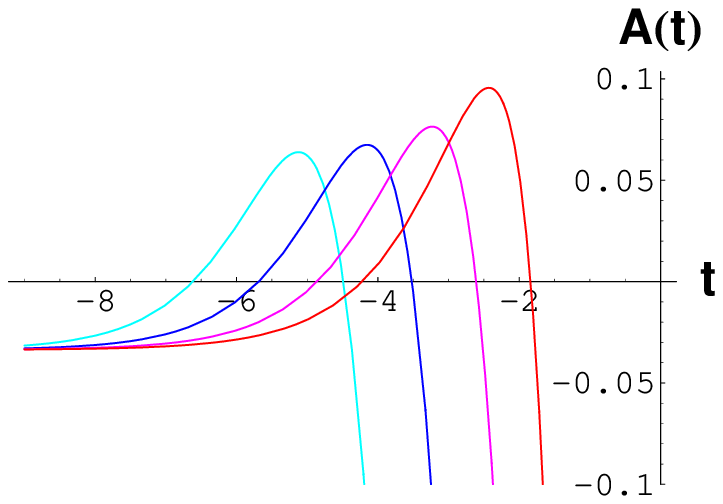}
\includegraphics[width=5cm]{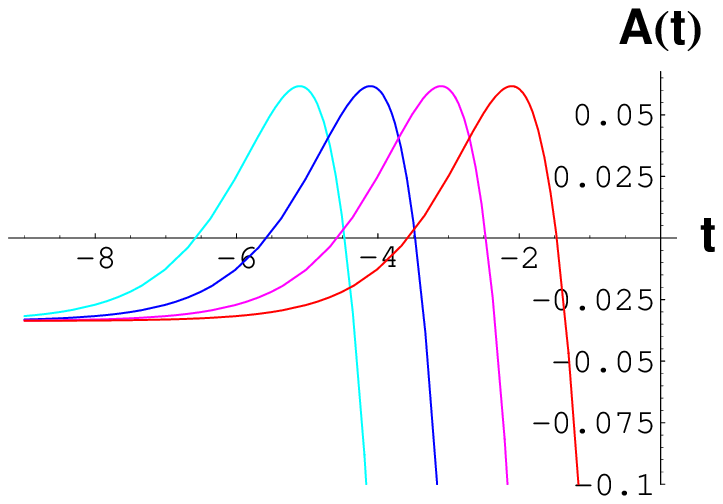}
\includegraphics[width=5cm]{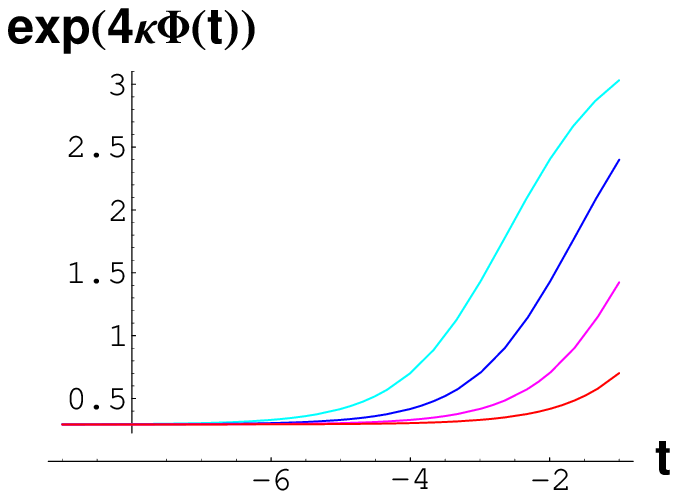}
\\
(a) \hspace{4cm} (b)  \hspace{4cm} (c)
\caption{(a) $A(t)$ for $k_b=-1$ as a function of $t$ in the canonical case. 
The curves correspond to the
cases with
$t_2=-3, -2, -1, 0$ and $t_2-t_3=1$, according to the location of the peak from
left to right. (b) $A(t)$ for $k_b=0$ as a function of $t$. 
The choice of parameters is the same
as (a). (c) $\exp(4\kappa\Phi(t))$ as a function of $t$.
The color of the curve corresponds to (a).
For the other parameters, see text.}
\label{fig2}
\end{figure}

We now consider the simplest case, $D=6$, $d=3$, $l=r=1$, $\alpha=1$, $q=1$ and
$t_1=0$.%
\footnote{Because the value of $q$ only determines the scale of the time coordinate
$t$, it does not concern the behavior of cosmic expansion.}
We show the values $A(t)\equiv-S(t)^{d-1}\frac{d^2S^{1-d}(t)}{dt^2}$ versus $t$ for
various values of $t_2$ and $t_3$ in Figs.~\ref{fig1} and \ref{fig2}.
It can be found that the acceleration period is earlier when $t_2-t_3$ is a
larger positive value. Contrarily, we find that negative $t_2-t_3$ leads to a
later period of acceleration.

We also find that if the value of $\alpha$ is sufficiently close to unity, there
exists an accelerating phase for a wide range of $t_2-t_1$ and $t_3-t_1$. We
conclude that the solution can give accelerating universes in the
$(d+1)$-dimensional Einstein frame in the model with $\sigma=1$, $l>0$, $r>0$ and
$k_b=-1$ and $k_b=0$. 
The result here is qualitatively similar to the result of Ref.~\cite{Roy},
where the dilaton coupling to a single flux term is considered.

\begin{figure}[ht]
\centering
\includegraphics[width=5cm]{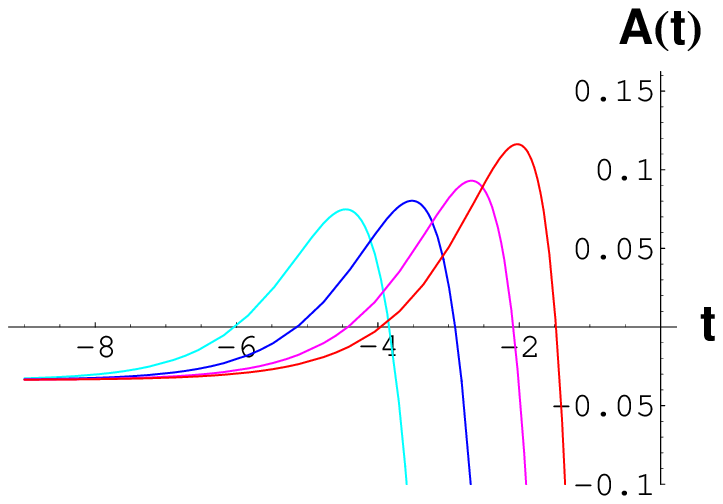}
\includegraphics[width=5cm]{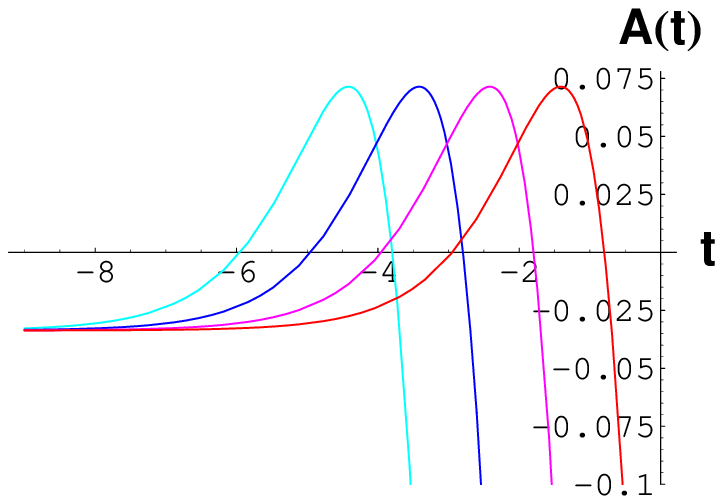}
\includegraphics[width=5cm]{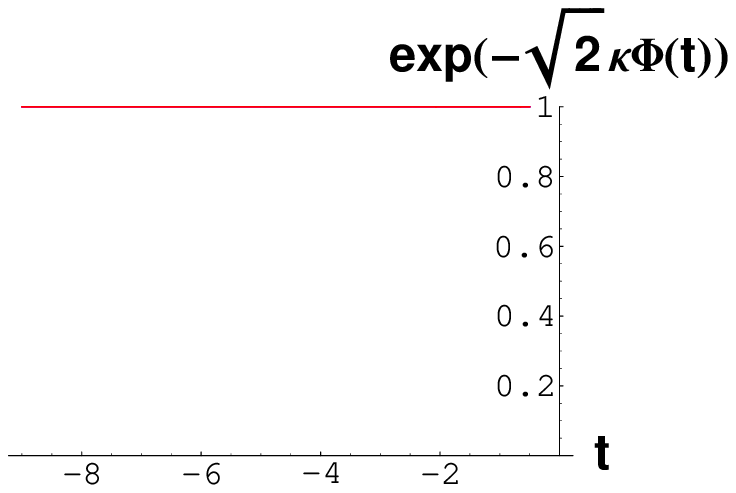}
\\
(a) \hspace{4cm} (b) \hspace{4cm} (c)
\caption{(a) $A(t)$ for $k_b=-1$ as a function of $t$ in the phantom case. 
The curves correspond to the
cases with
$t_2=t_3=-3, -2, -1, 0$, according to the location of the peak from left to right.
(b) $A(t)$ for $k_b=0$ as a function of $t$. The choice of parameters is the same
as (a). (c) $\exp(-\sqrt{2}\kappa\Phi(t))=1$ is constant in this case.
For the other parameters, see text.}
\label{fig3}
\end{figure}

\begin{figure}[ht]
\centering
\includegraphics[width=5cm]{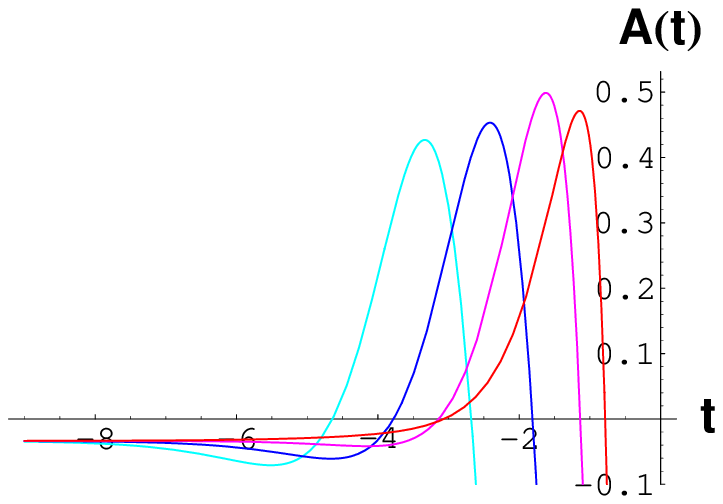}
\includegraphics[width=5cm]{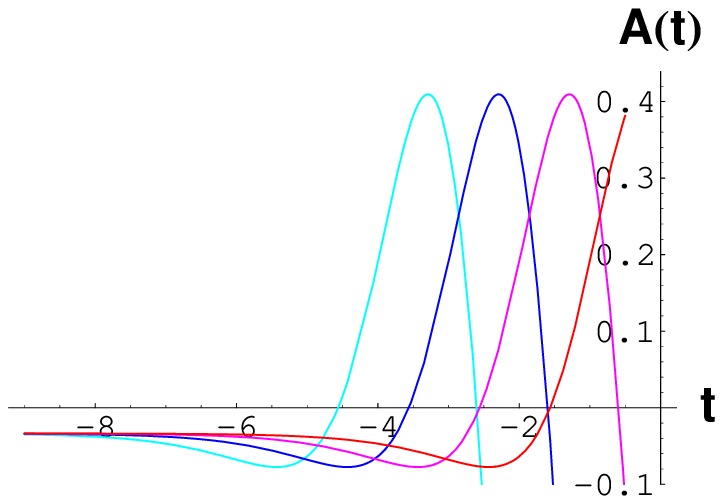}
\includegraphics[width=5cm]{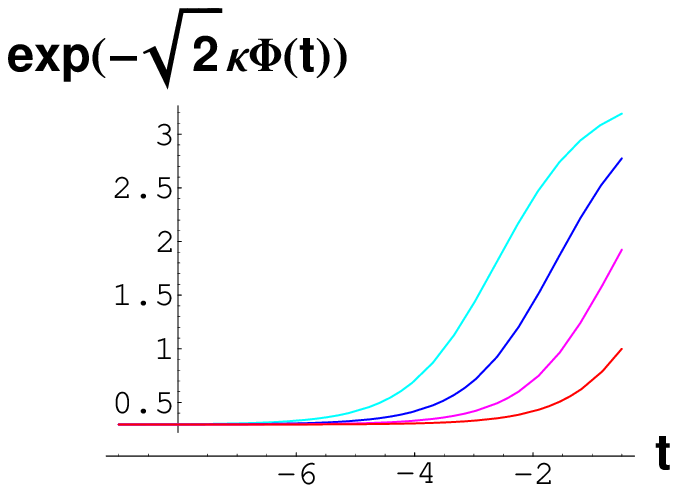}
\\
(a) \hspace{4cm} (b)  \hspace{4cm} (c)
\caption{(a) $A(t)$ for $k_b=-1$ as a function of $t$ in the phantom case. 
The curves correspond to the
cases with
$t_2=-3, -2, -1, 0$ and $t_2-t_3=1$, according to the location of the peak from
left to right. (b) $A(t)$ for $k_b=0$ as a function of $t$. 
The choice of parameters is the same
as (a). (c) $\exp(-\sqrt{2}\kappa\Phi(t))$ as a function of $t$.
The color of the curve corresponds to (a).
For the other parameters, see text.}
\label{fig4}
\end{figure}

A similar transient acceleration can be found in a special phantom case,
$\sigma=-1$,
$l>0$ and $r<0$, and ($E_1<0$, $E_2>0$, $E_3<0$).
In this case, we choose the integration constant to satisfy
$\lambda_3\sinh\theta=\lambda_2\cosh\theta$ in the solution (\ref{pp})
and obtain
\begin{equation}
e^{-2\kappa\left(\frac{1-\alpha^2}{\alpha}\right)\Phi(t)}=\frac{|r|}{l\alpha^2}
\frac{\cosh^2[q\kappa(t-t_3)]}{\cosh^2[q\kappa(t-t_2)]}\,.
\end{equation}
We now consider the simplest case, $D=6$, $d=3$, $l=-r=1$, $\alpha=1/\sqrt{2}$,
$q=1$ and $t_1=0$.
We show the values $A(t)$ versus $t$ for
various values of $t_2$ and $t_3$ in Figs.~\ref{fig3} and \ref{fig4} in this
phantom case. It can be found that the acceleration period is later when
$t_2-t_3$ is a larger positive value. Contrarily, negative $t_2-t_3$ leads to
an earlier period of acceleration.

\section{quantum cosmology}
\label{qc}
To study a very early universe and especially its initial state, we have to
consider the quantum nature of cosmology. In our model, we can obtain the
minisuperspace Wheeler--De\,Witt equation by replacing
$\dot{x}_a\rightarrow-i\frac{\partial}{\partial x_a}$ (where we choose the natural
unit $\hbar=1$) in the Hamiltonian $H$ and regarding the Hamiltonian constraint
as $H\Psi=0$, where $\Psi$ is the wave function of the universe.

Our Hamiltonian becomes
\begin{eqnarray}
& &H=\frac{1}{2}\frac{\partial^2}{\partial x^2}+\frac{V_1}{2}e^{2\lambda_1x}
-\frac{1}{2}\frac{\partial^2}{\partial y^2}+\frac{lf^2}{2}e^{2\lambda_2y}
-\frac{1}{2}\frac{\partial^2}{\partial z^2}+\frac{rf^2}{2}e^{2\lambda_3z}
\quad\mbox{for}\quad\sigma=1\,,\\
& &H=\frac{1}{2}\frac{\partial^2}{\partial x^2}+\frac{V_1}{2}e^{2\lambda_1x}
-\frac{1}{2}\frac{\partial^2}{\partial y^2}+\frac{lf^2}{2}e^{2\lambda_2y}
+\frac{1}{2}\frac{\partial^2}{\partial z^2}+\frac{rf^2}{2}e^{2\lambda_3z}
\quad\mbox{for}\quad\sigma=-1\,,
\end{eqnarray}
noticing that the definition of $\lambda_2$ and $\lambda_3$ is different in each
case.
Owing to the separation of variables,
the wave function is expressed by superposition of the 
multiplicative form,
$\Psi_1(x)\Psi_2(y)\Psi_3(z)$.

Let us first consider the case $\sigma=1$, $l>0$, and $r>0$.
Then, the normalizable wave function can be written in the form
\begin{eqnarray}
& &\Psi(x,y,z)=\int_{-\infty}^\infty dq\int_0^{2\pi}d\theta\, {\cal A}(q,\theta)
\left[c_1F_{i\frac{q}{\lambda_1}}({\sqrt{V_1}}e^{\lambda_1
x}/{\lambda_1})+
c_2G_{i\frac{q}{\lambda_1}}({\sqrt{V_1}}e^{\lambda_1
x}/{\lambda_1})\right]\nonumber \\
& &\qquad\qquad\qquad\qquad\times 
\frac{2(\sqrt{l}f/(2\lambda_2))^{-iq\cos\theta/\lambda_2}}%
{\Gamma\left(-iq\cos\theta/\lambda_2\right)}
K_{i\frac{q\cos\theta}{\lambda_2}}({\sqrt{l}f}e^{\lambda_2
y}/{\lambda_2})\nonumber \\
& &\qquad\qquad\qquad\qquad\times 
\frac{2(\sqrt{r}f/(2\lambda_3))^{-iq\sin\theta/\lambda_3}}%
{\Gamma\left(-iq\sin\theta/\lambda_3\right)}
K_{i\frac{q\sin\theta}{\lambda_3}}({\sqrt{r}f}e^{\lambda_3
z}/{\lambda_3})\,,
\label{wp1}
\end{eqnarray}
where ${\cal A}(q,\theta)$ is the amplitude and the eigenfunctions $F_\nu$ and
$G_\nu$ are defined as
\cite{ALNW,Dunster}
\begin{equation}
F_\nu(z)=\frac{1}{2\cos(\nu\pi/2)}[J_\nu(z)+J_{-\nu}(z)]\,,\quad
G_\nu(z)=\frac{1}{2\sin(\nu\pi/2)}[J_\nu(z)-J_{-\nu}(z)]\,,
\end{equation}
and $c_1$ and $c_2$ are constants.
In this expression, we adopt the wave normalization found in Ref.~\cite{LES1},
so
\begin{equation}
\frac{2(\sqrt{V}/(2\lambda))^{-iq/\lambda}}{\Gamma(-iq/\lambda)}
K_{iq/\lambda}(\sqrt{V}e^{\lambda x}/\lambda)\sim
e^{iqx}+R_0 e^{-iqx}\,\quad\mbox{for}\quad x\rightarrow -\infty\,,
\end{equation}
where
$R_0=[\Gamma(iq/\lambda)/\Gamma(-iq/\lambda)](\sqrt{V}/(2\lambda))^{-2iq/\lambda}$.

The Gaussian wave packet is often considered \cite{ALNW,KN,Kiefer1,Kiefer2,DKS} in
a semiclassical analysis of quantum cosmology.
There is another possibility that the amplitude ${\cal A}$ is independent of
$\theta$, which is naturally motivated from the form of (\ref{wp1})
because the integral region, or moduli space, of $\theta$ is apparently finite.

\begin{figure}[ht]
\centering
\includegraphics[width=7cm]{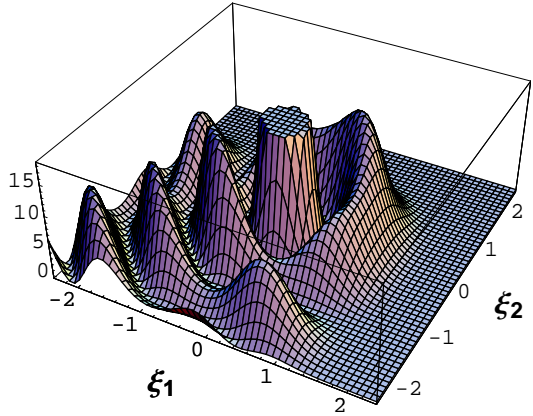}
\caption{$|\psi_q(\xi_1,\xi_2)|^2$ with $q=4$.}
\label{figqc}
\end{figure}

To succinctly grasp what occurs by taking this assumption,
we consider a simple calculation.
We show $|\psi_q(\xi_1,\xi_2)|^2$ with $q=4$ in Fig.~\ref{figqc}, where
\begin{equation}
\psi_q(\xi_1,\xi_2)\equiv\int_0^{2\pi}d\theta\,
\frac{2(2)^{iq\cos\theta}}{\Gamma(-iq\cos\theta)}
K_{i{q\cos\theta}}(e^{\xi_1})
\frac{2(2)^{iq\sin\theta}}{\Gamma(-iq\sin\theta)}
K_{i{q\sin\theta}}(e^{\xi_2})\,.
\end{equation}
Many peaks of the function are located in the region $(\xi_1<0,\xi_2<0)$,
and considerably high peaks are found at $\xi_1\sim\xi_2$. 
Because both the eigenfunction of $\xi_1$ and that of $\xi_2$ have the
incoming wave and reflected wave from the potential wall, the interference of the
four waves generates a complicated wave pattern.
Nevertheless, we find that a chain of peaks appears in the line
${\xi_1}\sim {\xi_2}$, which seems plausible from symmetry.

From this fact, we expect that, if the amplitude ${\cal A}(q)$
is assumed to have a sufficiently narrow width, peaks of the probability density
$|\Psi|^2$ appear at the discrete positions where
${\lambda_2y}\sim{\lambda_3z}$,
i.e., $\Phi\sim 0$ for $\alpha\sim 1$. 
From the shape of the partial wave function, we can see that the
initial state of the universe possesses a small but finite
$e^a$, whose possible value is somewhat discretized, with the stationary value
of $\Phi$.

Next, let us consider the phantom case $\sigma=-1$, $r<0$, and $k_b=1$.
Then, the normalizable wave function can be written in the form
\begin{eqnarray}
& &\Psi(x,y,z)=\int_{-\infty}^\infty dq\int_0^{2\pi}d\theta\, {\cal A}'(q,\theta)
\frac{2(\sqrt{V_1}/(2\lambda_1))^{-iq\cos\theta/\lambda_1}}%
{\Gamma(-iq\cos\theta/\lambda_1)}
K_{i\frac{q\cos\theta}{\lambda_1}}({\sqrt{|V_1|}}e^{\lambda_1
x}/{\lambda_1})\nonumber \\
& &\qquad\qquad\qquad\qquad\times
\left[c_1F_{i\frac{q}{\lambda_2}}({\sqrt{l}f}e^{\lambda_2
y}/{\lambda_2})+
c_2G_{i\frac{q}{\lambda_2}}({\sqrt{l}f}e^{\lambda_1
y}/{\lambda_2})\right] \nonumber \\
& &\qquad\qquad\qquad\qquad\times
\frac{2(\sqrt{|r|}f/(2\lambda_3))^{-iq\cos\theta/\lambda_3}}%
{\Gamma(-iq\cos\theta/\lambda_3)}
K_{i\frac{q\sin\theta}{\lambda_3}}({\sqrt{|r|}f}e^{\lambda_3
z}/{\lambda_3})\,,
\end{eqnarray}
where ${\cal A}'$ is the amplitude and we rearrange the integration constants.

We consider again the case ${\cal A}'={\cal A}'(q)$ as a Gaussian with a narrow
width. As in the previous case, we expect high peaks at
$\lambda_1x\sim\lambda_3z\lesssim 0$. Then, the initial state of the universe is
equipped with some finite $e^a$ and 
$b\propto \Phi$. 

The interpretation we present here is very qualitative, regrettably. To obtain
more quantitative results, we should study the wave function carefully, by taking
account of a normalization measure, which can be somewhat dependent on $\theta$,
and a detailed calculation of the superposition. This study will be done in future
work.
\section{Summary and discussion}
\label{sd}
In this paper, a class of analytical cosmological solutions is considered
in an integrable higher-dimensional model with a scalar field and an antisymmetric
tensor field. The scalar field is either a canonical ($\sigma=+1$) or a phantom
one  ($\sigma=-1$).

In Sec.~\ref{ac}, we looked for solutions for the accelerating universe.
We found that the expanding universe with transient acceleration is obtained 
in the case with hyperbolic or flat internal space with positive energy density of
the dilaton and antisymmetric fields ($\sigma=1$, $l>0$, and $r>0$), except for
other special cases.

One of the special features of our model is that the value of $\Phi$ can be finite
both at the beginning, $S\sim 0$, and at the far future of the universe,
$\eta\rightarrow\infty$. The coupling between the dilaton field and additional
gauge or matter fields can trigger some cosmological time-dependent phenomena,
though, unfortunately, it is difficult to interpret the present fields
in the
$(d+1)$-dimensional universe. The study of the inclusion of matter fields is
significant in any case, even apart from pursuing exact solutions.

We briefly argued the quantum cosmology of some specific cases.
Throughout this paper, we found that separable variables made the
analyses simple, both in the classical and the quantum cosmological behavior of the
scale factor. However, further deeper investigation is required, especially for
cosmology with initial bouncing behavior, which has not been discussed in the
present study.  We should also consider various aspects of wave functions of the
universe in our model in a more precise manner. We leave these subjects
for future work.

\appendix

\section{Summary of solutions}
\label{ss}
Here, we summarize the solutions and show the expressions for two scale factors
$a$ and $b$ and the scalar field $\Phi$. 
They are found by using $x$, $y$, and $z$, as follows:
\begin{eqnarray}
& &e^{2da(t)}=\left(e^{2\lambda_2y}\right)^{\frac{\sigma}{\alpha^2+\sigma}}
\left(e^{2\lambda_3z}\right)^{\frac{\alpha^2}{\alpha^2+\sigma}}\,,
\\&
&e^{2(D-d-2)b(t)}=e^{2\lambda_1x}\left(e^{2\lambda_2y}\right)^{-\frac{\sigma}{\alpha^2+\sigma}}
\left(e^{2\lambda_3z}\right)^{-\frac{\alpha^2}{\alpha^2+\sigma}}\,,
\\
& &e^{2\kappa
\Phi(t)}=\left(\frac{e^{2\lambda_2y}}{e^{2\lambda_3z}}
\right)^{\frac{\alpha}{\alpha^2+\sigma}}\,.
\end{eqnarray}

We exhibit them by categorizing the sign of $\sigma$, the signs of $l$ and $r$,
and the values for $k_b$, in this order.
In the following expressions, $t_1$, $t_2$, $t_3$, $q$, and $\theta$ are
integration constants.

{\scriptsize

\subsection{$\sigma=+1$}
%
\subsubsection{$l>0$ and $r>0$ ($E_1<0$, $E_2>0$, $E_3>0$)}
\begin{itemize}
\item $k_b=-1$
\begin{eqnarray}
& &e^{2da(t)}=\frac{1}{f^2}\left(\frac{q^2\cos^2\theta}{l\cosh^2[q\cos\theta
\lambda_2(t-t_2)]}
\right)^{\frac{1}{\alpha^2+1}}
\left(\frac{q^2\sin^2\theta}{r\cosh^2[q\sin\theta
\lambda_3(t-t_3)]}\right)^{\frac{\alpha^2}{\alpha^2+1}}\,,\nonumber \\
& &e^{2(D-d-2)b(t)}=\frac{f^2q^2}{V_1\sinh^2[q\lambda_1(t-t_1)]}
\left(\frac{q^2\cos^2\theta}{l\cosh^2[q\cos\theta
\lambda_2(t-t_2)]}
\right)^{-\frac{1}{\alpha^2+1}}
\left(\frac{q^2\sin^2\theta}{r\cosh^2[q\sin\theta
\lambda_3(t-t_3)]}\right)^{-\frac{\alpha^2}{\alpha^2+1}}\,,\nonumber \\
& &e^{2\kappa\Phi(t)}=\left(\frac{r\cos^2\theta}{l\sin^2\theta}
\frac{\cosh^2[q\sin\theta
\lambda_3(t-t_3)]}{\cosh^2[q\cos\theta
\lambda_2(t-t_2)]}
\right)^{\frac{\alpha}{\alpha^2+1}}
\,.\label{+++-;-++}
\end{eqnarray}
\item $k_b=0$
\begin{eqnarray}
& &e^{2da(t)}=\frac{1}{f^2}\left(\frac{q^2\cos^2\theta}{l\cosh^2[q\cos\theta
\lambda_2(t-t_2)]}
\right)^{\frac{1}{\alpha^2+1}}
\left(\frac{q^2\sin^2\theta}{r\cosh^2[q\sin\theta
\lambda_3(t-t_3)]}\right)^{\frac{\alpha^2}{\alpha^2+1}}\,,\nonumber \\
& &e^{2(D-d-2)b(t)}=C_1f^2e^{2\lambda_1qt}
\left(\frac{q^2\cos^2\theta}{l\cosh^2[q\cos\theta
\lambda_2(t-t_2)]}
\right)^{-\frac{1}{\alpha^2+1}}
\left(\frac{q^2\sin^2\theta}{r\cosh^2[q\sin\theta
\lambda_3(t-t_3)]}\right)^{-\frac{\alpha^2}{\alpha^2+1}}\,,\nonumber \\
& &e^{2\kappa\Phi(t)}=\left(\frac{r\cos^2\theta}{l\sin^2\theta}
\frac{\cosh^2[q\sin\theta
\lambda_3(t-t_3)]}{\cosh^2[q\cos\theta
\lambda_2(t-t_2)]}
\right)^{\frac{\alpha}{\alpha^2+1}}
\,,
\end{eqnarray}
where we define $C_1\equiv e^{-2q\lambda_1t_1}$. Hereafter, we use this
definition.
\item $k_b=+1$
\begin{eqnarray}
& &e^{2da(t)}=\frac{1}{f^2}\left(\frac{q^2\cos^2\theta}{l\cosh^2[q\cos\theta
\lambda_2(t-t_2)]}
\right)^{\frac{1}{\alpha^2+1}}
\left(\frac{q^2\sin^2\theta}{r\cosh^2[q\sin\theta
\lambda_3(t-t_3)]}\right)^{\frac{\alpha^2}{\alpha^2+1}}\,,\nonumber \\
& &e^{2(D-d-2)b(t)}=\frac{f^2q^2}{|V_1|\cosh^2[q\lambda_1(t-t_1)]}
\left(\frac{q^2\cos^2\theta}{l\cosh^2[q\cos\theta
\lambda_2(t-t_2)]}
\right)^{-\frac{1}{\alpha^2+1}}
\left(\frac{q^2\sin^2\theta}{r\cosh^2[q\sin\theta
\lambda_3(t-t_3)]}\right)^{-\frac{\alpha^2}{\alpha^2+1}}\,,\nonumber \\
& &e^{2\kappa\Phi(t)}=\left(\frac{r\cos^2\theta}{l\sin^2\theta}
\frac{\cosh^2[q\sin\theta
\lambda_3(t-t_3)]}{\cosh^2[q\cos\theta
\lambda_2(t-t_2)]}
\right)^{\frac{\alpha}{\alpha^2+1}}
\,.
\end{eqnarray}
\end{itemize}
\subsubsection{$l>0$ and $r<0$}
\begin{itemize}
\item $k_b=-1$
\begin{eqnarray}
\mbox{(I)}& &~(E_1<0, E_2>0, E_3>0)\nonumber \\
& &e^{2da(t)}=\frac{1}{f^2}\left(\frac{q^2\cos^2\theta}{l\cosh^2[q\cos\theta
\lambda_2(t-t_2)]}
\right)^{\frac{1}{\alpha^2+1}}
\left(\frac{q^2\sin^2\theta}{|r|\sinh^2[q\sin\theta
\lambda_3(t-t_3)]}\right)^{\frac{\alpha^2}{\alpha^2+1}}\,,\nonumber \\
& &e^{2(D-d-2)b(t)}=\frac{f^2q^2}{V_1\sinh^2[q\lambda_1(t-t_1)]}
\left(\frac{q^2\cos^2\theta}{l\cosh^2[q\cos\theta
\lambda_2(t-t_2)]}
\right)^{-\frac{1}{\alpha^2+1}}
\left(\frac{q^2\sin^2\theta}{|r|\sinh^2[q\sin\theta
\lambda_3(t-t_3)]}\right)^{-\frac{\alpha^2}{\alpha^2+1}}\,,\nonumber \\
& &e^{2\kappa\Phi(t)}=\left(\frac{|r|\cos^2\theta}{l\sin^2\theta}
\frac{\sinh^2[q\sin\theta
\lambda_3(t-t_3)]}{\cosh^2[q\cos\theta
\lambda_2(t-t_2)]}
\right)^{\frac{\alpha}{\alpha^2+1}}
\,.
\end{eqnarray}
\begin{eqnarray}
\mbox{(II)}& &~ (E_1<0, E_2>0, E_3<0)\nonumber \\
& &e^{2da(t)}=\frac{1}{f^2}\left(\frac{q^2\cosh^2\theta}{l\cosh^2[q\cosh\theta
\lambda_2(t-t_2)]}
\right)^{\frac{1}{\alpha^2+1}}
\left(\frac{q^2\sinh^2\theta}{|r|\sin^2[q\sinh\theta
\lambda_3(t-t_3)]}\right)^{\frac{\alpha^2}{\alpha^2+1}}\,,\nonumber \\
& &e^{2(D-d-2)b(t)}=\frac{f^2q^2}{V_1\sinh^2[q\lambda_1(t-t_1)]}
\left(\frac{q^2\cosh^2\theta}{l\cosh^2[q\cosh\theta
\lambda_2(t-t_2)]}
\right)^{-\frac{1}{\alpha^2+1}}
\left(\frac{q^2\sinh^2\theta}{|r|\sin^2[q\sinh\theta
\lambda_3(t-t_3)]}\right)^{-\frac{\alpha^2}{\alpha^2+1}}\,,\nonumber \\
& &e^{2\kappa\Phi(t)}=\left(\frac{|r|\cosh^2\theta}{l\sinh^2\theta}
\frac{\sin^2[q\sinh\theta
\lambda_3(t-t_3)]}{\cosh^2[q\cosh\theta
\lambda_2(t-t_2)]}
\right)^{\frac{\alpha}{\alpha^2+1}}
\,.
\end{eqnarray}
\begin{eqnarray}
\mbox{(III)}& &~(E_1>0, E_2>0, E_3<0)\nonumber \\
& &e^{2da(t)}=\frac{1}{f^2}\left(\frac{q^2\sinh^2\theta}{l\cosh^2[q\sinh\theta
\lambda_2(t-t_2)]}
\right)^{\frac{1}{\alpha^2+1}}
\left(\frac{q^2\cosh^2\theta}{|r|\sin^2[q\cosh\theta
\lambda_3(t-t_3)]}\right)^{\frac{\alpha^2}{\alpha^2+1}}\,,\nonumber \\
& &e^{2(D-d-2)b(t)}=\frac{f^2q^2}{V_1\sin^2[q\lambda_1(t-t_1)]}
\left(\frac{q^2\sinh^2\theta}{l\cosh^2[q\sinh\theta
\lambda_2(t-t_2)]}
\right)^{-\frac{1}{\alpha^2+1}}
\left(\frac{q^2\cosh^2\theta}{|r|\sin^2[q\cosh\theta
\lambda_3(t-t_3)]}\right)^{-\frac{\alpha^2}{\alpha^2+1}}\,,\nonumber \\
& &e^{2\kappa\Phi(t)}=\left(\frac{|r|\sinh^2\theta}{l\cosh^2\theta}
\frac{\sin^2[q\cosh\theta
\lambda_3(t-t_3)]}{\cosh^2[q\sinh\theta
\lambda_2(t-t_2)]}
\right)^{\frac{\alpha}{\alpha^2+1}}
\,.
\end{eqnarray}
\item $k_b=0$
\begin{eqnarray}
\mbox{(I)}& &~(E_1<0, E_2>0, E_3>0)\nonumber \\
& &e^{2da(t)}=\frac{1}{f^2}\left(\frac{q^2\cos^2\theta}{l\cosh^2[q\cos\theta
\lambda_2(t-t_2)]}
\right)^{\frac{1}{\alpha^2+1}}
\left(\frac{q^2\sin^2\theta}{|r|\sinh^2[q\sin\theta
\lambda_3(t-t_3)]}\right)^{\frac{\alpha^2}{\alpha^2+1}}\,,\nonumber \\
& &e^{2(D-d-2)b(t)}=C_1f^2e^{2\lambda_1qt}
\left(\frac{q^2\cos^2\theta}{l\cosh^2[q\cos\theta
\lambda_2(t-t_2)]}
\right)^{-\frac{1}{\alpha^2+1}}
\left(\frac{q^2\sin^2\theta}{|r|\sinh^2[q\sin\theta
\lambda_3(t-t_3)]}\right)^{-\frac{\alpha^2}{\alpha^2+1}}\,,\nonumber \\
& &e^{2\kappa\Phi(t)}=\left(\frac{|r|\cos^2\theta}{l\sin^2\theta}
\frac{\sinh^2[q\sin\theta
\lambda_3(t-t_3)]}{\cosh^2[q\cos\theta
\lambda_2(t-t_2)]}
\right)^{\frac{\alpha}{\alpha^2+1}}
\,.
\end{eqnarray}
\begin{eqnarray}
\mbox{(II)}& &~(E_1<0, E_2>0, E_3<0)\nonumber \\
& &e^{2da(t)}=\frac{1}{f^2}\left(\frac{q^2\cosh^2\theta}{l\cosh^2[q\cosh\theta
\lambda_2(t-t_2)]}
\right)^{\frac{1}{\alpha^2+1}}
\left(\frac{q^2\sinh^2\theta}{|r|\sin^2[q\sinh\theta
\lambda_3(t-t_3)]}\right)^{\frac{\alpha^2}{\alpha^2+1}}\,,\nonumber \\
& &e^{2(D-d-2)b(t)}=f^2e^{2\lambda_1q(t-t_1)}
\left(\frac{q^2\cosh^2\theta}{l\cosh^2[q\cosh\theta
\lambda_2(t-t_2)]}
\right)^{-\frac{1}{\alpha^2+1}}
\left(\frac{q^2\sinh^2\theta}{|r|\sin^2[q\sinh\theta
\lambda_3(t-t_3)]}\right)^{-\frac{\alpha^2}{\alpha^2+1}}\,,\nonumber \\
& &e^{2\kappa\Phi(t)}=\left(\frac{|r|\cosh^2\theta}{l\sinh^2\theta}
\frac{\sin^2[q\sinh\theta
\lambda_3(t-t_3)]}{\cosh^2[q\cosh\theta
\lambda_2(t-t_2)]}
\right)^{\frac{\alpha}{\alpha^2+1}}
\,.
\end{eqnarray}
\item $k_b=+1$
\begin{eqnarray}
\mbox{(I)}& &~(E_1<0, E_2>0, E_3>0)\nonumber \\
& &e^{2da(t)}=\frac{1}{f^2}\left(\frac{q^2\cos^2\theta}{l\cosh^2[q\cos\theta
\lambda_2(t-t_2)]}
\right)^{\frac{1}{\alpha^2+1}}
\left(\frac{q^2\sin^2\theta}{|r|\sinh^2[q\sin\theta
\lambda_3(t-t_3)]}\right)^{\frac{\alpha^2}{\alpha^2+1}}\,,\nonumber \\
& &e^{2(D-d-2)b(t)}=\frac{f^2q^2}{|V_1|\cosh^2[q\lambda_1(t-t_1)]}
\left(\frac{q^2\cos^2\theta}{l\cosh^2[q\cos\theta
\lambda_2(t-t_2)]}
\right)^{-\frac{1}{\alpha^2+1}}
\left(\frac{q^2\sin^2\theta}{|r|\sinh^2[q\sin\theta
\lambda_3(t-t_3)]}\right)^{-\frac{\alpha^2}{\alpha^2+1}}\,,\nonumber \\
& &e^{2\kappa\Phi(t)}=\left(\frac{|r|\cos^2\theta}{l\sin^2\theta}
\frac{\sinh^2[q\sin\theta
\lambda_3(t-t_3)]}{\cosh^2[q\cos\theta
\lambda_2(t-t_2)]}
\right)^{\frac{\alpha}{\alpha^2+1}}
\,.
\end{eqnarray}
\begin{eqnarray}
\mbox{(II)}& &~ (E_1<0, E_2>0, E_3<0)\nonumber \\
& &e^{2da(t)}=\frac{1}{f^2}\left(\frac{q^2\cosh^2\theta}{l\cosh^2[q\cosh\theta
\lambda_2(t-t_2)]}
\right)^{\frac{1}{\alpha^2+1}}
\left(\frac{q^2\sinh^2\theta}{|r|\sin^2[q\sinh\theta
\lambda_3(t-t_3)]}\right)^{\frac{\alpha^2}{\alpha^2+1}}\,,\nonumber \\
& &e^{2(D-d-2)b(t)}=\frac{f^2q^2}{|V_1|\cosh^2[q\lambda_1(t-t_1)]}
\left(\frac{q^2\cosh^2\theta}{l\cosh^2[q\cosh\theta
\lambda_2(t-t_2)]}
\right)^{-\frac{1}{\alpha^2+1}}
\left(\frac{q^2\sinh^2\theta}{|r|\sin^2[q\sinh\theta
\lambda_3(t-t_3)]}\right)^{-\frac{\alpha^2}{\alpha^2+1}}\,,\nonumber \\
& &e^{2\kappa\Phi(t)}=\left(\frac{|r|\cosh^2\theta}{l\sinh^2\theta}
\frac{\sin^2[q\sin\theta
\lambda_3(t-t_3)]}{\cosh^2[q\cosh\theta
\lambda_2(t-t_2)]}
\right)^{\frac{\alpha}{\alpha^2+1}}
\,.
\end{eqnarray}
\end{itemize}
\subsubsection{$l<0$ and $r>0$}
\begin{itemize}
\item $k_b=-1$
\begin{eqnarray}
\mbox{(I)}& &~ (E_1<0, E_2>0, E_3>0)\nonumber \\
& &e^{2da(t)}=\frac{1}{f^2}\left(\frac{q^2\cos^2\theta}{|l|\sinh^2[q\cos\theta
\lambda_2(t-t_2)]}
\right)^{\frac{1}{\alpha^2+1}}
\left(\frac{q^2\sin^2\theta}{r\cosh^2[q\sin\theta
\lambda_3(t-t_3)]}\right)^{\frac{\alpha^2}{\alpha^2+1}}\,,\nonumber \\
& &e^{2(D-d-2)b(t)}=\frac{f^2q^2}{V_1\sinh^2[q\lambda_1(t-t_1)]}
\left(\frac{q^2\cos^2\theta}{|l|\sinh^2[q\cos\theta
\lambda_2(t-t_2)]}
\right)^{-\frac{1}{\alpha^2+1}}
\left(\frac{q^2\sin^2\theta}{r\cosh^2[q\sin\theta
\lambda_3(t-t_3)]}\right)^{-\frac{\alpha^2}{\alpha^2+1}}\,,\nonumber \\
& &e^{2\kappa\Phi(t)}=\left(\frac{r\cos^2\theta}{|l|\sin^2\theta}
\frac{\cosh^2[q\sin\theta
\lambda_3(t-t_3)]}{\sinh^2[q\cos\theta
\lambda_2(t-t_2)]}
\right)^{\frac{\alpha}{\alpha^2+1}}
\,.
\end{eqnarray}
\begin{eqnarray}
\mbox{(II)}& &~ (E_1<0, E_2<0, E_3>0)\nonumber \\
& &e^{2da(t)}=\frac{1}{f^2}\left(\frac{q^2\sinh^2\theta}{|l|\sin^2[q\sinh\theta
\lambda_2(t-t_2)]}
\right)^{\frac{1}{\alpha^2+1}}
\left(\frac{q^2\cosh^2\theta}{r\cosh^2[q\cosh\theta
\lambda_3(t-t_3)]}\right)^{\frac{\alpha^2}{\alpha^2+1}}\,,\nonumber \\
& &e^{2(D-d-2)b(t)}=\frac{f^2q^2}{V_1\sinh^2[q\lambda_1(t-t_1)]}
\left(\frac{q^2\sinh^2\theta}{|l|\sin^2[q\sinh\theta
\lambda_2(t-t_2)]}
\right)^{-\frac{1}{\alpha^2+1}}
\left(\frac{q^2\cosh^2\theta}{r\cosh^2[q\cosh\theta
\lambda_3(t-t_3)]}\right)^{-\frac{\alpha^2}{\alpha^2+1}}\,,\nonumber \\
& &e^{2\kappa\Phi(t)}=\left(\frac{r\sinh^2\theta}{|l|\cosh^2\theta}
\frac{\cosh^2[q\cosh\theta
\lambda_3(t-t_3)]}{\sin^2[q\sinh\theta
\lambda_2(t-t_2)]}
\right)^{\frac{\alpha}{\alpha^2+1}}
\,.
\end{eqnarray}
\begin{eqnarray}
\mbox{(III)}& &~ (E_1>0, E_2<0, E_3>0)\nonumber \\
& &e^{2da(t)}=\frac{1}{f^2}\left(\frac{q^2\cosh^2\theta}{|l|\sin^2[q\cosh\theta
\lambda_2(t-t_2)]}
\right)^{\frac{1}{\alpha^2+1}}
\left(\frac{q^2\sinh^2\theta}{r\cosh^2[q\sinh\theta
\lambda_3(t-t_3)]}\right)^{\frac{\alpha^2}{\alpha^2+1}}\,,\nonumber \\
& &e^{2(D-d-2)b(t)}=\frac{f^2q^2}{V_1\sin^2[q\lambda_1(t-t_1)]}
\left(\frac{q^2\cosh^2\theta}{|l|\sin^2[q\cosh\theta
\lambda_2(t-t_2)]}
\right)^{-\frac{1}{\alpha^2+1}}
\left(\frac{q^2\sinh^2\theta}{r\cosh^2[q\sinh\theta
\lambda_3(t-t_3)]}\right)^{-\frac{\alpha^2}{\alpha^2+1}}\,,\nonumber \\
& &e^{2\kappa\Phi(t)}=\left(\frac{r\cosh^2\theta}{|l|\sinh^2\theta}
\frac{\cosh^2[q\sinh\theta
\lambda_3(t-t_3)]}{\sin^2[q\cosh\theta
\lambda_2(t-t_2)]}
\right)^{\frac{\alpha}{\alpha^2+1}}
\,.
\end{eqnarray}
\item $k_b=0$
\begin{eqnarray}
\mbox{(I)}& &~ (E_1<0, E_2>0, E_3>0)\nonumber \\
& &e^{2da(t)}=\frac{1}{f^2}\left(\frac{q^2\cos^2\theta}{|l|\sinh^2[q\cos\theta
\lambda_2(t-t_2)]}
\right)^{\frac{1}{\alpha^2+1}}
\left(\frac{q^2\sin^2\theta}{r\cosh^2[q\sin\theta
\lambda_3(t-t_3)]}\right)^{\frac{\alpha^2}{\alpha^2+1}}\,,\nonumber \\
& &e^{2(D-d-2)b(t)}=C_1f^2e^{2q\lambda_1t}
\left(\frac{q^2\cos^2\theta}{|l|\sinh^2[q\cos\theta
\lambda_2(t-t_2)]}
\right)^{-\frac{1}{\alpha^2+1}}
\left(\frac{q^2\sin^2\theta}{r\cosh^2[q\sin\theta
\lambda_3(t-t_3)]}\right)^{-\frac{\alpha^2}{\alpha^2+1}}\,,\nonumber \\
& &e^{2\kappa\Phi(t)}=\left(\frac{r\cos^2\theta}{|l|\sin^2\theta}
\frac{\cosh^2[q\sin\theta
\lambda_3(t-t_3)]}{\sinh^2[q\cos\theta
\lambda_2(t-t_2)]}
\right)^{\frac{\alpha}{\alpha^2+1}}
\,.
\end{eqnarray}
\begin{eqnarray}
\mbox{(II)}& &~ (E_1<0, E_2<0, E_3>0)\nonumber \\
& &e^{2da(t)}=\frac{1}{f^2}\left(\frac{q^2\sinh^2\theta}{|l|\sin^2[q\sinh\theta
\lambda_2(t-t_2)]}
\right)^{\frac{1}{\alpha^2+1}}
\left(\frac{q^2\cosh^2\theta}{r\cosh^2[q\cosh\theta
\lambda_3(t-t_3)]}\right)^{\frac{\alpha^2}{\alpha^2+1}}\,,\nonumber \\
& &e^{2(D-d-2)b(t)}=f^2e^{2q\lambda_1(t-t_1)}
\left(\frac{q^2\sinh^2\theta}{|l|\sin^2[q\sinh\theta
\lambda_2(t-t_2)]}
\right)^{-\frac{1}{\alpha^2+1}}
\left(\frac{q^2\cosh^2\theta}{r\cosh^2[q\cosh\theta
\lambda_3(t-t_3)]}\right)^{-\frac{\alpha^2}{\alpha^2+1}}\,,\nonumber \\
& &e^{2\kappa\Phi(t)}=\left(\frac{r\sinh^2\theta}{|l|\cosh^2\theta}
\frac{\cosh^2[q\cosh\theta
\lambda_3(t-t_3)]}{\sin^2[q\sinh\theta
\lambda_2(t-t_2)]}
\right)^{\frac{\alpha}{\alpha^2+1}}
\,.
\end{eqnarray}
\item $k_b=+1$
\begin{eqnarray}
\mbox{(I)}& &~ (E_1<0, E_2>0, E_3>0)\nonumber \\
& &e^{2da(t)}=\frac{1}{f^2}\left(\frac{q^2\cos^2\theta}{|l|\sinh^2[q\cos\theta
\lambda_2(t-t_2)]}
\right)^{\frac{1}{\alpha^2+1}}
\left(\frac{q^2\sin^2\theta}{r\cosh^2[q\sin\theta
\lambda_3(t-t_3)]}\right)^{\frac{\alpha^2}{\alpha^2+1}}\,,\nonumber \\
& &e^{2(D-d-2)b(t)}=\frac{f^2q^2}{|V_1|\cosh^2[q\lambda_1(t-t_1)]}
\left(\frac{q^2\cos^2\theta}{|l|\sinh^2[q\cos\theta
\lambda_2(t-t_2)]}
\right)^{-\frac{1}{\alpha^2+1}}
\left(\frac{q^2\sin^2\theta}{r\cosh^2[q\sin\theta
\lambda_3(t-t_3)]}\right)^{-\frac{\alpha^2}{\alpha^2+1}}\,,\nonumber \\
& &e^{2\kappa\Phi(t)}=\left(\frac{r\cos^2\theta}{|l|\sin^2\theta}
\frac{\cosh^2[q\sin\theta
\lambda_3(t-t_3)]}{\sinh^2[q\cos\theta
\lambda_2(t-t_2)]}
\right)^{\frac{\alpha}{\alpha^2+1}}
\,.
\end{eqnarray}
\begin{eqnarray}
\mbox{(II)}& &~ (E_1<0, E_2<0, E_3>0)\nonumber \\
& &e^{2da(t)}=\frac{1}{f^2}\left(\frac{q^2\sinh^2\theta}{|l|\sin^2[q\sinh\theta
\lambda_2(t-t_2)]}
\right)^{\frac{1}{\alpha^2+1}}
\left(\frac{q^2\cosh^2\theta}{r\cosh^2[q\cosh\theta
\lambda_3(t-t_3)]}\right)^{\frac{\alpha^2}{\alpha^2+1}}\,,\nonumber \\
& &e^{2(D-d-2)b(t)}=\frac{f^2q^2}{|V_1|\cosh^2[q\lambda_1(t-t_1)]}
\left(\frac{q^2\sinh^2\theta}{|l|\sin^2[q\sinh\theta
\lambda_2(t-t_2)]}
\right)^{-\frac{1}{\alpha^2+1}}
\left(\frac{q^2\cosh^2\theta}{r\cosh^2[q\cosh\theta
\lambda_3(t-t_3)]}\right)^{-\frac{\alpha^2}{\alpha^2+1}}\,,\nonumber \\
& &e^{2\kappa\Phi(t)}=\left(\frac{r\sinh^2\theta}{|l|\cosh^2\theta}
\frac{\cosh^2[q\cosh\theta
\lambda_3(t-t_3)]}{\sin^2[q\sinh\theta
\lambda_2(t-t_2)]}
\right)^{\frac{\alpha}{\alpha^2+1}}
\,.
\end{eqnarray}
\end{itemize}
\subsubsection{$l<0$ and $r<0$}
\begin{itemize}
\item $k_b=-1$
\begin{eqnarray}
\mbox{(I)}& &~ (E_1<0, E_2>0, E_3>0)\nonumber \\
& &e^{2da(t)}=\frac{1}{f^2}\left(\frac{q^2\cos^2\theta}{|l|\sinh^2[q\cos\theta
\lambda_2(t-t_2)]}
\right)^{\frac{1}{\alpha^2+1}}
\left(\frac{q^2\sin^2\theta}{|r|\sinh^2[q\sin\theta
\lambda_3(t-t_3)]}\right)^{\frac{\alpha^2}{\alpha^2+1}}\,,\nonumber \\
& &e^{2(D-d-2)b(t)}=\frac{f^2q^2}{V_1\sinh^2[q\lambda_1(t-t_1)]}
\left(\frac{q^2\cos^2\theta}{|l|\sinh^2[q\cos\theta
\lambda_2(t-t_2)]}
\right)^{-\frac{1}{\alpha^2+1}}
\left(\frac{q^2\sin^2\theta}{|r|\sinh^2[q\sin\theta
\lambda_3(t-t_3)]}\right)^{-\frac{\alpha^2}{\alpha^2+1}}\,,\nonumber \\
& &e^{2\kappa\Phi(t)}=\left(\frac{|r|\cos^2\theta}{|l|\sin^2\theta}
\frac{\sinh^2[q\sin\theta
\lambda_3(t-t_3)]}{\sinh^2[q\cos\theta
\lambda_2(t-t_2)]}
\right)^{\frac{\alpha}{\alpha^2+1}}
\,.\label{+---;-++}
\end{eqnarray}
\begin{eqnarray}
\mbox{(II)}& &~ (E_1<0, E_2>0, E_3<0)\nonumber \\
& &e^{2da(t)}=\frac{1}{f^2}\left(\frac{q^2\cosh^2\theta}{|l|\sinh^2[q\cosh\theta
\lambda_2(t-t_2)]}
\right)^{\frac{1}{\alpha^2+1}}
\left(\frac{q^2\sinh^2\theta}{|r|\sin^2[q\sinh\theta
\lambda_3(t-t_3)]}\right)^{\frac{\alpha^2}{\alpha^2+1}}\,,\nonumber \\
& &e^{2(D-d-2)b(t)}=\frac{f^2q^2}{V_1\sinh^2[q\lambda_1(t-t_1)]}
\left(\frac{q^2\cosh^2\theta}{|l|\sinh^2[q\cosh\theta
\lambda_2(t-t_2)]}
\right)^{-\frac{1}{\alpha^2+1}}
\left(\frac{q^2\sinh^2\theta}{|r|\sin^2[q\sinh\theta
\lambda_3(t-t_3)]}\right)^{-\frac{\alpha^2}{\alpha^2+1}}\,,\nonumber \\
& &e^{2\kappa\Phi(t)}=\left(\frac{|r|\cosh^2\theta}{|l|\sinh^2\theta}
\frac{\sin^2[q\sinh\theta
\lambda_3(t-t_3)]}{\sinh^2[q\cosh\theta
\lambda_2(t-t_2)]}
\right)^{\frac{\alpha}{\alpha^2+1}}
\,.\label{+---;-+-}
\end{eqnarray}
\begin{eqnarray}
\mbox{(III)}& &~ (E_1<0, E_2<0, E_3>0)\nonumber \\
& &e^{2da(t)}=\frac{1}{f^2}\left(\frac{q^2\sinh^2\theta}{|l|\sin^2[q\sinh\theta
\lambda_2(t-t_2)]}
\right)^{\frac{1}{\alpha^2+1}}
\left(\frac{q^2\cosh^2\theta}{|r|\sinh^2[q\cosh\theta
\lambda_3(t-t_3)]}\right)^{\frac{\alpha^2}{\alpha^2+1}}\,,\nonumber \\
& &e^{2(D-d-2)b(t)}=\frac{f^2q^2}{V_1\sinh^2[q\lambda_1(t-t_1)]}
\left(\frac{q^2\sinh^2\theta}{|l|\sin^2[q\sinh\theta
\lambda_2(t-t_2)]}
\right)^{-\frac{1}{\alpha^2+1}}
\left(\frac{q^2\cosh^2\theta}{|r|\sinh^2[q\cosh\theta
\lambda_3(t-t_3)]}\right)^{-\frac{\alpha^2}{\alpha^2+1}}\,,\nonumber \\
& &e^{2\kappa\Phi(t)}=\left(\frac{|r|\sinh^2\theta}{|l|\cosh^2\theta}
\frac{\sinh^2[q\cosh\theta
\lambda_3(t-t_3)]}{\sin^2[q\sinh\theta
\lambda_2(t-t_2)]}
\right)^{\frac{\alpha}{\alpha^2+1}}
\,.\label{+---;--+}
\end{eqnarray}
\begin{eqnarray}
\mbox{(IV)}& &~ (E_1>0, E_2<0, E_3<0)\nonumber \\
& &e^{2da(t)}=\frac{1}{f^2}\left(\frac{q^2\cos^2\theta}{|l|\sin^2[q\cos\theta
\lambda_2(t-t_2)]}
\right)^{\frac{1}{\alpha^2+1}}
\left(\frac{q^2\sin^2\theta}{|r|\sin^2[q\sin\theta
\lambda_3(t-t_3)]}\right)^{\frac{\alpha^2}{\alpha^2+1}}\,,\nonumber \\
& &e^{2(D-d-2)b(t)}=\frac{f^2q^2}{V_1\sin^2[q\lambda_1(t-t_1)]}
\left(\frac{q^2\cos^2\theta}{|l|\sin^2[q\cos\theta
\lambda_2(t-t_2)]}
\right)^{-\frac{1}{\alpha^2+1}}
\left(\frac{q^2\sin^2\theta}{|r|\sin^2[q\sin\theta
\lambda_3(t-t_3)]}\right)^{-\frac{\alpha^2}{\alpha^2+1}}\,,\nonumber \\
& &e^{2\kappa\Phi(t)}=\left(\frac{|r|\cos^2\theta}{|l|\sin^2\theta}
\frac{\sin^2[q\sin\theta
\lambda_3(t-t_3)]}{\sin^2[q\cos\theta
\lambda_2(t-t_2)]}
\right)^{\frac{\alpha}{\alpha^2+1}}
\,.\label{+---;+--}
\end{eqnarray}
\begin{eqnarray}
\mbox{(V)}& &~ (E_1>0, E_2>0, E_3<0)\nonumber \\
& &e^{2da(t)}=\frac{1}{f^2}\left(\frac{q^2\sinh^2\theta}{|l|\sinh^2[q\sinh\theta
\lambda_2(t-t_2)]}
\right)^{\frac{1}{\alpha^2+1}}
\left(\frac{q^2\cosh^2\theta}{|r|\sin^2[q\cosh\theta
\lambda_3(t-t_3)]}\right)^{\frac{\alpha^2}{\alpha^2+1}}\,,\nonumber \\
& &e^{2(D-d-2)b(t)}=\frac{f^2q^2}{V_1\sin^2[q\lambda_1(t-t_1)]}
\left(\frac{q^2\sinh^2\theta}{|l|\sinh^2[q\sinh\theta
\lambda_2(t-t_2)]}
\right)^{-\frac{1}{\alpha^2+1}}
\left(\frac{q^2\cosh^2\theta}{|r|\sin^2[q\cosh\theta
\lambda_3(t-t_3)]}\right)^{-\frac{\alpha^2}{\alpha^2+1}}\,,\nonumber \\
& &e^{2\kappa\Phi(t)}=\left(\frac{|r|\sinh^2\theta}{|l|\cosh^2\theta}
\frac{\sin^2[q\cosh\theta
\lambda_3(t-t_3)]}{\sinh^2[q\sinh\theta
\lambda_2(t-t_2)]}
\right)^{\frac{\alpha}{\alpha^2+1}}
\,.\label{+---;++-}
\end{eqnarray}
\begin{eqnarray}
\mbox{(VI)}& &~ (E_1>0, E_2<0, E_3>0)\nonumber \\
& &e^{2da(t)}=\frac{1}{f^2}\left(\frac{q^2\cosh^2\theta}{|l|\sin^2[q\cosh\theta
\lambda_2(t-t_2)]}
\right)^{\frac{1}{\alpha^2+1}}
\left(\frac{q^2\sinh^2\theta}{|r|\sinh^2[q\sinh\theta
\lambda_3(t-t_3)]}\right)^{\frac{\alpha^2}{\alpha^2+1}}\,,\nonumber \\
& &e^{2(D-d-2)b(t)}=\frac{f^2q^2}{V_1\sin^2[q\lambda_1(t-t_1)]}
\left(\frac{q^2\cosh^2\theta}{|l|\sin^2[q\cosh\theta
\lambda_2(t-t_2)]}
\right)^{-\frac{1}{\alpha^2+1}}
\left(\frac{q^2\sinh^2\theta}{|r|\sinh^2[q\sinh\theta
\lambda_3(t-t_3)]}\right)^{-\frac{\alpha^2}{\alpha^2+1}}\,,\nonumber \\
& &e^{2\kappa\Phi(t)}=\left(\frac{|r|\cosh^2\theta}{|l|\sinh^2\theta}
\frac{\sinh^2[q\sinh\theta
\lambda_3(t-t_3)]}{\sin^2[q\cosh\theta
\lambda_2(t-t_2)]}
\right)^{\frac{\alpha}{\alpha^2+1}}
\,.\label{+---;+-+}
\end{eqnarray}
\item $k_b=0$
\begin{eqnarray}
\mbox{(I)}& &~ (E_1<0, E_2>0, E_3>0)\nonumber \\
& &e^{2da(t)}=\frac{1}{f^2}\left(\frac{q^2\cos^2\theta}{|l|\sinh^2[q\cos\theta
\lambda_2(t-t_2)]}
\right)^{\frac{1}{\alpha^2+1}}
\left(\frac{q^2\sin^2\theta}{|r|\sinh^2[q\sin\theta
\lambda_3(t-t_3)]}\right)^{\frac{\alpha^2}{\alpha^2+1}}\,,\nonumber \\
& &e^{2(D-d-2)b(t)}=C_1f^2e^{2q\lambda_1t}
\left(\frac{q^2\cos^2\theta}{|l|\sinh^2[q\cos\theta
\lambda_2(t-t_2)]}
\right)^{-\frac{1}{\alpha^2+1}}
\left(\frac{q^2\sin^2\theta}{|r|\sinh^2[q\sin\theta
\lambda_3(t-t_3)]}\right)^{-\frac{\alpha^2}{\alpha^2+1}}\,,\nonumber \\
& &e^{2\kappa\Phi(t)}=\left(\frac{|r|\cos^2\theta}{|l|\sin^2\theta}
\frac{\sinh^2[q\sin\theta
\lambda_3(t-t_3)]}{\sinh^2[q\cos\theta
\lambda_2(t-t_2)]}
\right)^{\frac{\alpha}{\alpha^2+1}}
\,.\label{+--0;-++}
\end{eqnarray}
\begin{eqnarray}
\mbox{(II)}& &~ (E_1<0, E_2>0, E_3<0)\nonumber \\
& &e^{2da(t)}=\frac{1}{f^2}\left(\frac{q^2\cosh^2\theta}{|l|\sinh^2[q\cosh\theta
\lambda_2(t-t_2)]}
\right)^{\frac{1}{\alpha^2+1}}
\left(\frac{q^2\sinh^2\theta}{|r|\sin^2[q\sinh\theta
\lambda_3(t-t_3)]}\right)^{\frac{\alpha^2}{\alpha^2+1}}\,,\nonumber \\
& &e^{2(D-d-2)b(t)}=f^2e^{2q\lambda_1(t-t_1)}
\left(\frac{q^2\cosh^2\theta}{|l|\sinh^2[q\cosh\theta
\lambda_2(t-t_2)]}
\right)^{-\frac{1}{\alpha^2+1}}
\left(\frac{q^2\sinh^2\theta}{|r|\sin^2[q\sinh\theta
\lambda_3(t-t_3)]}\right)^{-\frac{\alpha^2}{\alpha^2+1}}\,,\nonumber \\
& &e^{2\kappa\Phi(t)}=\left(\frac{|r|\cosh^2\theta}{|l|\sinh^2\theta}
\frac{\sin^2[q\sinh\theta
\lambda_3(t-t_3)]}{\sinh^2[q\cosh\theta
\lambda_2(t-t_2)]}
\right)^{\frac{\alpha}{\alpha^2+1}}
\,.\label{+--0;-+-}
\end{eqnarray}
\begin{eqnarray}
\mbox{(III)}& &~ (E_1<0, E_2<0, E_3>0)\nonumber \\
& &e^{2da(t)}=\frac{1}{f^2}\left(\frac{q^2\sinh^2\theta}{|l|\sin^2[q\sinh\theta
\lambda_2(t-t_2)]}
\right)^{\frac{1}{\alpha^2+1}}
\left(\frac{q^2\cosh^2\theta}{|r|\sinh^2[q\cosh\theta
\lambda_3(t-t_3)]}\right)^{\frac{\alpha^2}{\alpha^2+1}}\,,\nonumber \\
& &e^{2(D-d-2)b(t)}=f^2e^{2q\lambda_1(t-t_1)}
\left(\frac{q^2\sinh^2\theta}{|l|\sin^2[q\sinh\theta
\lambda_2(t-t_2)]}
\right)^{-\frac{1}{\alpha^2+1}}
\left(\frac{q^2\cosh^2\theta}{|r|\sinh^2[q\cosh\theta
\lambda_3(t-t_3)]}\right)^{-\frac{\alpha^2}{\alpha^2+1}}\,,\nonumber \\
& &e^{2\kappa\Phi(t)}=\left(\frac{|r|\sinh^2\theta}{|l|\cosh^2\theta}
\frac{\sinh^2[q\cosh\theta
\lambda_3(t-t_3)]}{\sin^2[q\sinh\theta
\lambda_2(t-t_2)]}
\right)^{\frac{\alpha}{\alpha^2+1}}
\,.\label{+--0;--+}
\end{eqnarray}
\item $k_b=+1$
\begin{eqnarray}
\mbox{(I)}& &~ (E_1<0, E_2>0, E_3>0)\nonumber \\
& &e^{2da(t)}=\frac{1}{f^2}\left(\frac{q^2\cos^2\theta}{|l|\sinh^2[q\cos\theta
\lambda_2(t-t_2)]}
\right)^{\frac{1}{\alpha^2+1}}
\left(\frac{q^2\sin^2\theta}{|r|\sinh^2[q\sin\theta
\lambda_3(t-t_3)]}\right)^{\frac{\alpha^2}{\alpha^2+1}}\,,\nonumber \\
& &e^{2(D-d-2)b(t)}=\frac{f^2q^2}{|V_1|\cosh^2[q\lambda_1(t-t_1)]}
\left(\frac{q^2\cos^2\theta}{|l|\sinh^2[q\cos\theta
\lambda_2(t-t_2)]}
\right)^{-\frac{1}{\alpha^2+1}}
\left(\frac{q^2\sin^2\theta}{|r|\sinh^2[q\sin\theta
\lambda_3(t-t_3)]}\right)^{-\frac{\alpha^2}{\alpha^2+1}}\,,\nonumber \\
& &e^{2\kappa\Phi(t)}=\left(\frac{|r|\cos^2\theta}{|l|\sin^2\theta}
\frac{\sinh^2[q\sin\theta
\lambda_3(t-t_3)]}{\sinh^2[q\cos\theta
\lambda_2(t-t_2)]}
\right)^{\frac{\alpha}{\alpha^2+1}}
\,,
\end{eqnarray}
\begin{eqnarray}
\mbox{(II)}& &~ (E_1<0, E_2>0, E_3<0)\nonumber \\
& &e^{2da(t)}=\frac{1}{f^2}\left(\frac{q^2\cosh^2\theta}{|l|\sinh^2[q\cosh\theta
\lambda_2(t-t_2)]}
\right)^{\frac{1}{\alpha^2+1}}
\left(\frac{q^2\sinh^2\theta}{|r|\sin^2[q\sinh\theta
\lambda_3(t-t_3)]}\right)^{\frac{\alpha^2}{\alpha^2+1}}\,,\nonumber \\
& &e^{2(D-d-2)b(t)}=\frac{f^2q^2}{|V_1|\cosh^2[q\lambda_1(t-t_1)]}
\left(\frac{q^2\cosh^2\theta}{|l|\sinh^2[q\cosh\theta
\lambda_2(t-t_2)]}
\right)^{-\frac{1}{\alpha^2+1}}
\left(\frac{q^2\sinh^2\theta}{|r|\sin^2[q\sinh\theta
\lambda_3(t-t_3)]}\right)^{-\frac{\alpha^2}{\alpha^2+1}}\,,\nonumber \\
& &e^{2\kappa\Phi(t)}=\left(\frac{|r|\cosh^2\theta}{|l|\sinh^2\theta}
\frac{\sin^2[q\sinh\theta
\lambda_3(t-t_3)]}{\sinh^2[q\cosh\theta
\lambda_2(t-t_2)]}
\right)^{\frac{\alpha}{\alpha^2+1}}
\,.
\end{eqnarray}
\begin{eqnarray}
\mbox{(III)}& &~ (E_1<0, E_2<0, E_3>0)\nonumber \\
& &e^{2da(t)}=\frac{1}{f^2}\left(\frac{q^2\sinh^2\theta}{|l|\sin^2[q\sinh\theta
\lambda_2(t-t_2)]}
\right)^{\frac{1}{\alpha^2+1}}
\left(\frac{q^2\cosh^2\theta}{|r|\sinh^2[q\cosh\theta
\lambda_3(t-t_3)]}\right)^{\frac{\alpha^2}{\alpha^2+1}}\,,\nonumber \\
& &e^{2(D-d-2)b(t)}=\frac{f^2q^2}{|V_1|\cosh^2[q\lambda_1(t-t_1)]}
\left(\frac{q^2\sinh^2\theta}{|l|\sin^2[q\sinh\theta
\lambda_2(t-t_2)]}
\right)^{-\frac{1}{\alpha^2+1}}
\left(\frac{q^2\cosh^2\theta}{|r|\sinh^2[q\cosh\theta
\lambda_3(t-t_3)]}\right)^{-\frac{\alpha^2}{\alpha^2+1}}\,,\nonumber \\
& &e^{2\kappa\Phi(t)}=\left(\frac{|r|\sinh^2\theta}{|l|\cosh^2\theta}
\frac{\sinh^2[q\cosh\theta
\lambda_3(t-t_3)]}{\sin^2[q\sinh\theta
\lambda_2(t-t_2)]}
\right)^{\frac{\alpha}{\alpha^2+1}}
\,.
\end{eqnarray}
\end{itemize}
\subsection{$\sigma=-1$}
%
\subsubsection{$l>0$ and $r>0$}
\begin{itemize}
\item $k_b=-1$
\begin{eqnarray}
\mbox{(I)}& &~ (E_1<0, E_2>0, E_3>0)\nonumber \\
& &e^{2da(t)}=\frac{1}{f^2}\left(\frac{q^2\cos^2\theta}{l\cosh^2[q\cos\theta
\lambda_2(t-t_2)]}
\right)^{\frac{1}{1-\alpha^2}}
\left(\frac{q^2\sin^2\theta}{r\sin^2[q\sin\theta
\lambda_3(t-t_3)]}\right)^{-\frac{\alpha^2}{1-\alpha^2}}\,,\nonumber \\
& &e^{2(D-d-2)b(t)}=\frac{f^2q^2}{V_1\sinh^2[q\lambda_1(t-t_1)]}
\left(\frac{q^2\cos^2\theta}{l\cosh^2[q\cos\theta
\lambda_2(t-t_2)]}
\right)^{-\frac{1}{1-\alpha^2}}
\left(\frac{q^2\sin^2\theta}{r\sin^2[q\sin\theta
\lambda_3(t-t_3)]}\right)^{\frac{\alpha^2}{1-\alpha^2}}\,,\nonumber \\
& &e^{2\kappa\Phi(t)}=\left(\frac{r\cos^2\theta}{l\sin^2\theta}
\frac{\sin^2[q\sin\theta
\lambda_3(t-t_3)]}{\cosh^2[q\cos\theta
\lambda_2(t-t_2)]}
\right)^{-\frac{\alpha}{1-\alpha^2}}
\,.
\end{eqnarray}
\begin{eqnarray}
\mbox{(II)}& &~ (E_1<0, E_2>0, E_3<0)\nonumber \\
& &e^{2da(t)}=\frac{1}{f^2}\left(\frac{q^2\cosh^2\theta}{l\cosh^2[q\cosh\theta
\lambda_2(t-t_2)]}
\right)^{\frac{1}{1-\alpha^2}}
\left(\frac{q^2\sinh^2\theta}{r\sinh^2[q\sinh\theta
\lambda_3(t-t_3)]}\right)^{-\frac{\alpha^2}{1-\alpha^2}}\,,\nonumber \\
& &e^{2(D-d-2)b(t)}=\frac{f^2q^2}{V_1\sinh^2[q\lambda_1(t-t_1)]}
\left(\frac{q^2\cosh^2\theta}{l\cosh^2[q\cosh\theta
\lambda_2(t-t_2)]}
\right)^{-\frac{1}{1-\alpha^2}}
\left(\frac{q^2\sinh^2\theta}{r\sinh^2[q\sinh\theta
\lambda_3(t-t_3)]}\right)^{\frac{\alpha^2}{1-\alpha^2}}\,,\nonumber \\
& &e^{2\kappa\Phi(t)}=\left(\frac{r\cosh^2\theta}{l\sinh^2\theta}
\frac{\sinh^2[q\sinh\theta
\lambda_3(t-t_3)]}{\cosh^2[q\cosh\theta
\lambda_2(t-t_2)]}
\right)^{-\frac{\alpha}{1-\alpha^2}}
\,.
\end{eqnarray}
\begin{eqnarray}
\mbox{(III)}& &~(E_1>0, E_2>0, E_3<0)\nonumber \\
& &e^{2da(t)}=\frac{1}{f^2}\left(\frac{q^2\sinh^2\theta}{l\cosh^2[q\sinh\theta
\lambda_2(t-t_2)]}
\right)^{\frac{1}{1-\alpha^2}}
\left(\frac{q^2\cosh^2\theta}{r\sinh^2[q\cosh\theta
\lambda_3(t-t_3)]}\right)^{-\frac{\alpha^2}{1-\alpha^2}}\,,\nonumber \\
& &e^{2(D-d-2)b(t)}=\frac{f^2q^2}{V_1\sin^2[q\lambda_1(t-t_1)]}
\left(\frac{q^2\sinh^2\theta}{l\cosh^2[q\sinh\theta
\lambda_2(t-t_2)]}
\right)^{-\frac{1}{1-\alpha^2}}
\left(\frac{q^2\cosh^2\theta}{r\sinh^2[q\cosh\theta
\lambda_3(t-t_3)]}\right)^{\frac{\alpha^2}{1-\alpha^2}}\,,\nonumber \\
& &e^{2\kappa\Phi(t)}=\left(\frac{r\sinh^2\theta}{l\cosh^2\theta}
\frac{\sinh^2[q\cosh\theta
\lambda_3(t-t_3)]}{\cosh^2[q\sinh\theta
\lambda_2(t-t_2)]}
\right)^{-\frac{\alpha}{1-\alpha^2}}
\,.
\end{eqnarray}
\item $k_b=0$
\begin{eqnarray}
\mbox{(I)}& &~ (E_1<0, E_2>0, E_3>0)\nonumber \\
& &e^{2da(t)}=\frac{1}{f^2}\left(\frac{q^2\cos^2\theta}{l\cosh^2[q\cos\theta
\lambda_2(t-t_2)]}
\right)^{\frac{1}{1-\alpha^2}}
\left(\frac{q^2\sin^2\theta}{r\sin^2[q\sin\theta
\lambda_3(t-t_3)]}\right)^{-\frac{\alpha^2}{1-\alpha^2}}\,,\nonumber \\
& &e^{2(D-d-2)b(t)}=C_1f^2e^{2q\lambda_1t}
\left(\frac{q^2\cos^2\theta}{l\cosh^2[q\cos\theta
\lambda_2(t-t_2)]}
\right)^{-\frac{1}{1-\alpha^2}}
\left(\frac{q^2\sin^2\theta}{r\sin^2[q\sin\theta
\lambda_3(t-t_3)]}\right)^{\frac{\alpha^2}{1-\alpha^2}}\,,\nonumber \\
& &e^{2\kappa\Phi(t)}=\left(\frac{r\cos^2\theta}{l\sin^2\theta}
\frac{\sin^2[q\sin\theta
\lambda_3(t-t_3)]}{\cosh^2[q\cos\theta
\lambda_2(t-t_2)]}
\right)^{-\frac{\alpha}{1-\alpha^2}}
\,.
\end{eqnarray}
\begin{eqnarray}
\mbox{(II)}& &~ (E_1<0, E_2>0, E_3<0)\nonumber \\
& &e^{2da(t)}=\frac{1}{f^2}\left(\frac{q^2\cosh^2\theta}{l\cosh^2[q\cosh\theta
\lambda_2(t-t_2)]}
\right)^{\frac{1}{1-\alpha^2}}
\left(\frac{q^2\sinh^2\theta}{r\sinh^2[q\sinh\theta
\lambda_3(t-t_3)]}\right)^{-\frac{\alpha^2}{1-\alpha^2}}\,,\nonumber \\
& &e^{2(D-d-2)b(t)}=f^2e^{2q\lambda_1(t-t_1)}
\left(\frac{q^2\cosh^2\theta}{l\cosh^2[q\cosh\theta
\lambda_2(t-t_2)]}
\right)^{-\frac{1}{1-\alpha^2}}
\left(\frac{q^2\sinh^2\theta}{r\sinh^2[q\sinh\theta
\lambda_3(t-t_3)]}\right)^{\frac{\alpha^2}{1-\alpha^2}}\,,\nonumber \\
& &e^{2\kappa\Phi(t)}=\left(\frac{r\cosh^2\theta}{l\sinh^2\theta}
\frac{\sinh^2[q\sinh\theta
\lambda_3(t-t_3)]}{\cosh^2[q\cosh\theta
\lambda_2(t-t_2)]}
\right)^{-\frac{\alpha}{1-\alpha^2}}
\,.
\end{eqnarray}
\item $k_b=+1$
\begin{eqnarray}
\mbox{(I)}& &~ (E_1<0, E_2>0, E_3>0)\nonumber \\
& &e^{2da(t)}=\frac{1}{f^2}\left(\frac{q^2\cos^2\theta}{l\cosh^2[q\cos\theta
\lambda_2(t-t_2)]}
\right)^{\frac{1}{1-\alpha^2}}
\left(\frac{q^2\sin^2\theta}{r\sin^2[q\sin\theta
\lambda_3(t-t_3)]}\right)^{-\frac{\alpha^2}{1-\alpha^2}}\,,\nonumber \\
& &e^{2(D-d-2)b(t)}=\frac{f^2q^2}{|V_1|\cosh^2[q\lambda_1(t-t_1)]}
\left(\frac{q^2\cos^2\theta}{l\cosh^2[q\cos\theta
\lambda_2(t-t_2)]}
\right)^{-\frac{1}{1-\alpha^2}}
\left(\frac{q^2\sin^2\theta}{r\sin^2[q\sin\theta
\lambda_3(t-t_3)]}\right)^{\frac{\alpha^2}{1-\alpha^2}}\,,\nonumber \\
& &e^{2\kappa\Phi(t)}=\left(\frac{r\cos^2\theta}{l\sin^2\theta}
\frac{\sin^2[q\sin\theta
\lambda_3(t-t_3)]}{\cosh^2[q\cos\theta
\lambda_2(t-t_2)]}
\right)^{-\frac{\alpha}{1-\alpha^2}}
\,.
\end{eqnarray}
\begin{eqnarray}
\mbox{(II)}& &~ (E_1<0, E_2>0, E_3<0)\nonumber \\
& &e^{2da(t)}=\frac{1}{f^2}\left(\frac{q^2\cosh^2\theta}{l\cosh^2[q\cosh\theta
\lambda_2(t-t_2)]}
\right)^{\frac{1}{1-\alpha^2}}
\left(\frac{q^2\sinh^2\theta}{r\sinh^2[q\sinh\theta
\lambda_3(t-t_3)]}\right)^{-\frac{\alpha^2}{1-\alpha^2}}\,,\nonumber \\
& &e^{2(D-d-2)b(t)}=\frac{f^2q^2}{|V_1|\cosh^2[q\lambda_1(t-t_1)]}
\left(\frac{q^2\cosh^2\theta}{l\cosh^2[q\cosh\theta
\lambda_2(t-t_2)]}
\right)^{-\frac{1}{1-\alpha^2}}
\left(\frac{q^2\sinh^2\theta}{r\sinh^2[q\sinh\theta
\lambda_3(t-t_3)]}\right)^{\frac{\alpha^2}{1-\alpha^2}}\,,\nonumber \\
& &e^{2\kappa\Phi(t)}=\left(\frac{r\cosh^2\theta}{l\sinh^2\theta}
\frac{\sinh^2[q\sinh\theta
\lambda_3(t-t_3)]}{\cosh^2[q\cosh\theta
\lambda_2(t-t_2)]}
\right)^{-\frac{\alpha}{1-\alpha^2}}
\,.
\end{eqnarray}
\end{itemize}
\subsubsection{$l>0$ and $r<0$}
\begin{itemize}
\item $k_b=-1$
\begin{eqnarray}
\mbox{(I)}& &~ (E_1<0, E_2>0, E_3<0)\nonumber \\
& &e^{2da(t)}=\frac{1}{f^2}\left(\frac{q^2\cosh^2\theta}{l\cosh^2[q\cosh\theta
\lambda_2(t-t_2)]}
\right)^{\frac{1}{1-\alpha^2}}
\left(\frac{q^2\sinh^2\theta}{|r|\cosh^2[q\sinh\theta
\lambda_3(t-t_3)]}\right)^{-\frac{\alpha^2}{1-\alpha^2}}\,,\nonumber \\
& &e^{2(D-d-2)b(t)}=\frac{f^2q^2}{V_1\sinh^2[q\lambda_1(t-t_1)]}
\left(\frac{q^2\cosh^2\theta}{l\cosh^2[q\cosh\theta
\lambda_2(t-t_2)]}
\right)^{-\frac{1}{1-\alpha^2}}
\left(\frac{q^2\sinh^2\theta}{|r|\cosh^2[q\sinh\theta
\lambda_3(t-t_3)]}\right)^{\frac{\alpha^2}{1-\alpha^2}}\,,\nonumber \\
& &e^{2\kappa\Phi(t)}=\left(\frac{|r|\cosh^2\theta}{l\sinh^2\theta}
\frac{\cosh^2[q\sinh\theta
\lambda_3(t-t_3)]}{\cosh^2[q\cosh\theta
\lambda_2(t-t_2)]}
\right)^{-\frac{\alpha}{1-\alpha^2}}
\,.\label{pp}
\end{eqnarray}
\begin{eqnarray}
\mbox{(II)}& &~ (E_1>0, E_2>0, E_3<0)\nonumber \\
& &e^{2da(t)}=\frac{1}{f^2}\left(\frac{q^2\sinh^2\theta}{l\cosh^2[q\sinh\theta
\lambda_2(t-t_2)]}
\right)^{\frac{1}{1-\alpha^2}}
\left(\frac{q^2\cosh^2\theta}{|r|\cosh^2[q\cosh\theta
\lambda_3(t-t_3)]}\right)^{-\frac{\alpha^2}{1-\alpha^2}}\,,\nonumber \\
& &e^{2(D-d-2)b(t)}=\frac{f^2q^2}{V_1\sin^2[q\lambda_1(t-t_1)]}
\left(\frac{q^2\sinh^2\theta}{l\cosh^2[q\sinh\theta
\lambda_2(t-t_2)]}
\right)^{-\frac{1}{1-\alpha^2}}
\left(\frac{q^2\cosh^2\theta}{|r|\cosh^2[q\cosh\theta
\lambda_3(t-t_3)]}\right)^{\frac{\alpha^2}{1-\alpha^2}}\,,\nonumber \\
& &e^{2\kappa\Phi(t)}=\left(\frac{|r|\sinh^2\theta}{l\cosh^2\theta}
\frac{\cosh^2[q\cosh\theta
\lambda_3(t-t_3)]}{\cosh^2[q\sinh\theta
\lambda_2(t-t_2)]}
\right)^{-\frac{\alpha}{1-\alpha^2}}
\,.
\end{eqnarray}
\item $k_b=0$ ($E_1<0$, $E_2>0$, $E_3<0$)
\begin{eqnarray}
& &e^{2da(t)}=\frac{1}{f^2}\left(\frac{q^2\cosh^2\theta}{l\cosh^2[q\cosh\theta
\lambda_2(t-t_2)]}
\right)^{\frac{1}{1-\alpha^2}}
\left(\frac{q^2\sinh^2\theta}{|r|\cosh^2[q\sinh\theta
\lambda_3(t-t_3)]}\right)^{-\frac{\alpha^2}{1-\alpha^2}}\,,\nonumber \\
& &e^{2(D-d-2)b(t)}=C_1f^2e^{2q\lambda_1t}
\left(\frac{q^2\cosh^2\theta}{l\cosh^2[q\cosh\theta
\lambda_2(t-t_2)]}
\right)^{-\frac{1}{1-\alpha^2}}
\left(\frac{q^2\sinh^2\theta}{|r|\cosh^2[q\sinh\theta
\lambda_3(t-t_3)]}\right)^{\frac{\alpha^2}{1-\alpha^2}}\,,\nonumber \\
& &e^{2\kappa\Phi(t)}=\left(\frac{|r|\cosh^2\theta}{l\sinh^2\theta}
\frac{\cosh^2[q\sinh\theta
\lambda_3(t-t_3)]}{\cosh^2[q\cosh\theta
\lambda_2(t-t_2)]}
\right)^{-\frac{\alpha}{1-\alpha^2}}
\,.
\end{eqnarray}
\item $k_b=+1$ ($E_1<0$, $E_2>0$, $E_3<0$)
\begin{eqnarray}
& &e^{2da(t)}=\frac{1}{f^2}\left(\frac{q^2\cosh^2\theta}{l\cosh^2[q\cosh\theta
\lambda_2(t-t_2)]}
\right)^{\frac{1}{1-\alpha^2}}
\left(\frac{q^2\sinh^2\theta}{|r|\cosh^2[q\sinh\theta
\lambda_3(t-t_3)]}\right)^{-\frac{\alpha^2}{1-\alpha^2}}\,,\nonumber \\
& &e^{2(D-d-2)b(t)}=\frac{f^2q^2}{|V_1|\cosh^2[q\lambda_1(t-t_1)]}
\left(\frac{q^2\cosh^2\theta}{l\cosh^2[q\cosh\theta
\lambda_2(t-t_2)]}
\right)^{-\frac{1}{1-\alpha^2}}
\left(\frac{q^2\sinh^2\theta}{|r|\cosh^2[q\sinh\theta
\lambda_3(t-t_3)]}\right)^{\frac{\alpha^2}{1-\alpha^2}}\,,\nonumber \\
& &e^{2\kappa\Phi(t)}=\left(\frac{|r|\cosh^2\theta}{l\sinh^2\theta}
\frac{\cosh^2[q\sinh\theta
\lambda_3(t-t_3)]}{\cosh^2[q\cosh\theta
\lambda_2(t-t_2)]}
\right)^{-\frac{\alpha}{1-\alpha^2}}
\,.
\end{eqnarray}
\end{itemize}
\subsubsection{$l<0$ and $r>0$}
\begin{itemize}
\item $k_b=-1$
\begin{eqnarray}
\mbox{(I)}& &~ (E_1<0, E_2>0, E_3>0)\nonumber \\
& &e^{2da(t)}=\frac{1}{f^2}\left(\frac{q^2\cos^2\theta}{|l|\sinh^2[q\cos\theta
\lambda_2(t-t_2)]}
\right)^{\frac{1}{1-\alpha^2}}
\left(\frac{q^2\sin^2\theta}{r\sin^2[q\sin\theta
\lambda_3(t-t_3)]}\right)^{-\frac{\alpha^2}{1-\alpha^2}}\,,\nonumber \\
& &e^{2(D-d-2)b(t)}=\frac{f^2q^2}{V_1\sinh^2[q\lambda_1(t-t_1)]}
\left(\frac{q^2\cos^2\theta}{|l|\sinh^2[q\cos\theta
\lambda_2(t-t_2)]}
\right)^{-\frac{1}{1-\alpha^2}}
\left(\frac{q^2\sin^2\theta}{r\sin^2[q\sin\theta
\lambda_3(t-t_3)]}\right)^{\frac{\alpha^2}{1-\alpha^2}}\,,\nonumber \\
& &e^{2\kappa\Phi(t)}=\left(\frac{r\cos^2\theta}{|l|\sin^2\theta}
\frac{\sin^2[q\sin\theta
\lambda_3(t-t_3)]}{\sinh^2[q\cos\theta
\lambda_2(t-t_2)]}
\right)^{-\frac{\alpha}{1-\alpha^2}}
\,.\label{--+-;-++}
\end{eqnarray}
\begin{eqnarray}
\mbox{(II)}& &~ (E_1<0, E_2>0, E_3<0)\nonumber \\
& &e^{2da(t)}=\frac{1}{f^2}\left(\frac{q^2\cosh^2\theta}{|l|\sinh^2[q\cosh\theta
\lambda_2(t-t_2)]}
\right)^{\frac{1}{1-\alpha^2}}
\left(\frac{q^2\sinh^2\theta}{r\sinh^2[q\sinh\theta
\lambda_3(t-t_3)]}\right)^{-\frac{\alpha^2}{1-\alpha^2}}\,,\nonumber \\
& &e^{2(D-d-2)b(t)}=\frac{f^2q^2}{V_1\sinh^2[q\lambda_1(t-t_1)]}
\left(\frac{q^2\cosh^2\theta}{|l|\sinh^2[q\cosh\theta
\lambda_2(t-t_2)]}
\right)^{-\frac{1}{1-\alpha^2}}
\left(\frac{q^2\sinh^2\theta}{r\sinh^2[q\sinh\theta
\lambda_3(t-t_3)]}\right)^{\frac{\alpha^2}{1-\alpha^2}}\,,\nonumber \\
& &e^{2\kappa\Phi(t)}=\left(\frac{r\cosh^2\theta}{|l|\sinh^2\theta}
\frac{\sinh^2[q\sinh\theta
\lambda_3(t-t_3)]}{\sinh^2[q\cosh\theta
\lambda_2(t-t_2)]}
\right)^{-\frac{\alpha}{1-\alpha^2}}
\,.\label{--+-;-+-}
\end{eqnarray}
\begin{eqnarray}
\mbox{(III)}& &~ (E_1<0, E_2<0, E_3>0)\nonumber \\
& &e^{2da(t)}=\frac{1}{f^2}\left(\frac{q^2\sinh^2\theta}{|l|\sin^2[q\sinh\theta
\lambda_2(t-t_2)]}
\right)^{\frac{1}{1-\alpha^2}}
\left(\frac{q^2\cosh^2\theta}{r\sin^2[q\cosh\theta
\lambda_3(t-t_3)]}\right)^{-\frac{\alpha^2}{1-\alpha^2}}\,,\nonumber \\
& &e^{2(D-d-2)b(t)}=\frac{f^2q^2}{V_1\sinh^2[q\lambda_1(t-t_1)]}
\left(\frac{q^2\sinh^2\theta}{|l|\sin^2[q\sinh\theta
\lambda_2(t-t_2)]}
\right)^{-\frac{1}{1-\alpha^2}}
\left(\frac{q^2\cosh^2\theta}{r\sin^2[q\cosh\theta
\lambda_3(t-t_3)]}\right)^{\frac{\alpha^2}{1-\alpha^2}}\,,\nonumber \\
& &e^{2\kappa\Phi(t)}=\left(\frac{r\sinh^2\theta}{|l|\cosh^2\theta}
\frac{\sin^2[q\cosh\theta
\lambda_3(t-t_3)]}{\sin^2[q\sinh\theta
\lambda_2(t-t_2)]}
\right)^{-\frac{\alpha}{1-\alpha^2}}
\,.\label{--+-;--+}
\end{eqnarray}
\begin{eqnarray}
\mbox{(IV)}& &~ (E_1>0, E_2>0, E_3<0)\nonumber \\
& &e^{2da(t)}=\frac{1}{f^2}\left(\frac{q^2\sinh^2\theta}{|l|\sinh^2[q\sinh\theta
\lambda_2(t-t_2)]}
\right)^{\frac{1}{1-\alpha^2}}
\left(\frac{q^2\cosh^2\theta}{r\sinh^2[q\cosh\theta
\lambda_3(t-t_3)]}\right)^{-\frac{\alpha^2}{1-\alpha^2}}\,,\nonumber \\
& &e^{2(D-d-2)b(t)}=\frac{f^2q^2}{V_1\sin^2[q\lambda_1(t-t_1)]}
\left(\frac{q^2\sinh^2\theta}{|l|\sinh^2[q\sinh\theta
\lambda_2(t-t_2)]}
\right)^{-\frac{1}{1-\alpha^2}}
\left(\frac{q^2\cosh^2\theta}{r\sinh^2[q\cosh\theta
\lambda_3(t-t_3)]}\right)^{\frac{\alpha^2}{1-\alpha^2}}\,,\nonumber \\
& &e^{2\kappa\Phi(t)}=\left(\frac{r\sinh^2\theta}{|l|\cosh^2\theta}
\frac{\sinh^2[q\cosh\theta
\lambda_3(t-t_3)]}{\sinh^2[q\sinh\theta
\lambda_2(t-t_2)]}
\right)^{-\frac{\alpha}{1-\alpha^2}}
\,.\label{--+-;++-}
\end{eqnarray}
\begin{eqnarray}
\mbox{(V)}& &~ (E_1>0, E_2<0, E_3>0)\nonumber \\
& &e^{2da(t)}=\frac{1}{f^2}\left(\frac{q^2\cosh^2\theta}{|l|\sin^2[q\cosh\theta
\lambda_2(t-t_2)]}
\right)^{\frac{1}{1-\alpha^2}}
\left(\frac{q^2\sinh^2\theta}{r\sin^2[q\sinh\theta
\lambda_3(t-t_3)]}\right)^{-\frac{\alpha^2}{1-\alpha^2}}\,,\nonumber \\
& &e^{2(D-d-2)b(t)}=\frac{f^2q^2}{V_1\sin^2[q\lambda_1(t-t_1)]}
\left(\frac{q^2\cosh^2\theta}{|l|\sin^2[q\cosh\theta
\lambda_2(t-t_2)]}
\right)^{-\frac{1}{1-\alpha^2}}
\left(\frac{q^2\sinh^2\theta}{r\sin^2[q\sinh\theta
\lambda_3(t-t_3)]}\right)^{\frac{\alpha^2}{1-\alpha^2}}\,,\nonumber \\
& &e^{2\kappa\Phi(t)}=\left(\frac{r\cosh^2\theta}{|l|\sinh^2\theta}
\frac{\sin^2[q\sinh\theta
\lambda_3(t-t_3)]}{\sin^2[q\cosh\theta
\lambda_2(t-t_2)]}
\right)^{-\frac{\alpha}{1-\alpha^2}}
\,.\label{--+-;+-+}
\end{eqnarray}
\begin{eqnarray}
\mbox{(VI)}& &~ (E_1>0, E_2<0, E_3<0)\nonumber \\
& &e^{2da(t)}=\frac{1}{f^2}\left(\frac{q^2\cos^2\theta}{|l|\sin^2[q\cos\theta
\lambda_2(t-t_2)]}
\right)^{\frac{1}{1-\alpha^2}}
\left(\frac{q^2\sin^2\theta}{r\sinh^2[q\sin\theta
\lambda_3(t-t_3)]}\right)^{-\frac{\alpha^2}{1-\alpha^2}}\,,\nonumber \\
& &e^{2(D-d-2)b(t)}=\frac{f^2q^2}{V_1\sin^2[q\lambda_1(t-t_1)]}
\left(\frac{q^2\cos^2\theta}{|l|\sin^2[q\cos\theta
\lambda_2(t-t_2)]}
\right)^{-\frac{1}{1-\alpha^2}}
\left(\frac{q^2\sin^2\theta}{r\sinh^2[q\sin\theta
\lambda_3(t-t_3)]}\right)^{\frac{\alpha^2}{1-\alpha^2}}\,,\nonumber \\
& &e^{2\kappa\Phi(t)}=\left(\frac{r\cos^2\theta}{|l|\sin^2\theta}
\frac{\sinh^2[q\sin\theta
\lambda_3(t-t_3)]}{\sin^2[q\cos\theta
\lambda_2(t-t_2)]}
\right)^{-\frac{\alpha}{1-\alpha^2}}
\,.\label{--+-;+--}
\end{eqnarray}
\item $k_b=0$
\begin{eqnarray}
\mbox{(I)}& &~ (E_1<0, E_2>0, E_3>0)\nonumber \\
& &e^{2da(t)}=\frac{1}{f^2}\left(\frac{q^2\cos^2\theta}{|l|\sinh^2[q\cos\theta
\lambda_2(t-t_2)]}
\right)^{\frac{1}{1-\alpha^2}}
\left(\frac{q^2\sin^2\theta}{r\sin^2[q\sin\theta
\lambda_3(t-t_3)]}\right)^{-\frac{\alpha^2}{1-\alpha^2}}\,,\nonumber \\
& &e^{2(D-d-2)b(t)}=C_1f^2e^{2q\lambda_1t}
\left(\frac{q^2\cos^2\theta}{|l|\sinh^2[q\cos\theta
\lambda_2(t-t_2)]}
\right)^{-\frac{1}{1-\alpha^2}}
\left(\frac{q^2\sin^2\theta}{r\sin^2[q\sin\theta
\lambda_3(t-t_3)]}\right)^{\frac{\alpha^2}{1-\alpha^2}}\,,\nonumber \\
& &e^{2\kappa\Phi(t)}=\left(\frac{r\cos^2\theta}{|l|\sin^2\theta}
\frac{\sin^2[q\sin\theta
\lambda_3(t-t_3)]}{\sinh^2[q\cos\theta
\lambda_2(t-t_2)]}
\right)^{-\frac{\alpha}{1-\alpha^2}}
\,.\label{--+0;-++}
\end{eqnarray}
\begin{eqnarray}
\mbox{(II)}& &~ (E_1<0, E_2>0, E_3<0)\nonumber \\
& &e^{2da(t)}=\frac{1}{f^2}\left(\frac{q^2\cosh^2\theta}{|l|\sinh^2[q\cosh\theta
\lambda_2(t-t_2)]}
\right)^{\frac{1}{1-\alpha^2}}
\left(\frac{q^2\sinh^2\theta}{r\sinh^2[q\sinh\theta
\lambda_3(t-t_3)]}\right)^{-\frac{\alpha^2}{1-\alpha^2}}\,,\nonumber \\
& &e^{2(D-d-2)b(t)}=f^2e^{2q\lambda_1(t-t_1)}
\left(\frac{q^2\cosh^2\theta}{|l|\sinh^2[q\cosh\theta
\lambda_2(t-t_2)]}
\right)^{-\frac{1}{1-\alpha^2}}
\left(\frac{q^2\sinh^2\theta}{r\sinh^2[q\sinh\theta
\lambda_3(t-t_3)]}\right)^{\frac{\alpha^2}{1-\alpha^2}}\,,\nonumber \\
& &e^{2\kappa\Phi(t)}=\left(\frac{r\cosh^2\theta}{|l|\sinh^2\theta}
\frac{\sinh^2[q\sinh\theta
\lambda_3(t-t_3)]}{\sinh^2[q\cosh\theta
\lambda_2(t-t_2)]}
\right)^{-\frac{\alpha}{1-\alpha^2}}
\,.\label{--+0;-+-}
\end{eqnarray}
\begin{eqnarray}
\mbox{(III)}& &~ (E_1<0, E_2<0, E_3>0)\nonumber \\
& &e^{2da(t)}=\frac{1}{f^2}\left(\frac{q^2\sinh^2\theta}{|l|\sin^2[q\sinh\theta
\lambda_2(t-t_2)]}
\right)^{\frac{1}{1-\alpha^2}}
\left(\frac{q^2\cosh^2\theta}{r\sin^2[q\cosh\theta
\lambda_3(t-t_3)]}\right)^{-\frac{\alpha^2}{1-\alpha^2}}\,,\nonumber \\
& &e^{2(D-d-2)b(t)}=f^2e^{2q\lambda_1(t-t_1)}
\left(\frac{q^2\sin^2\theta}{|l|\sin^2[q\sinh\theta
\lambda_2(t-t_2)]}
\right)^{-\frac{1}{1-\alpha^2}}
\left(\frac{q^2\cosh^2\theta}{r\sin^2[q\cosh\theta
\lambda_3(t-t_3)]}\right)^{\frac{\alpha^2}{1-\alpha^2}}\,,\nonumber \\
& &e^{2\kappa\Phi(t)}=\left(\frac{r\sinh^2\theta}{|l|\cosh^2\theta}
\frac{\sin^2[q\cosh\theta
\lambda_3(t-t_3)]}{\sin^2[q\sinh\theta
\lambda_2(t-t_2)]}
\right)^{-\frac{\alpha}{1-\alpha^2}}
\,.\label{--+0;--+}
\end{eqnarray}
\item $k_b=+1$
\begin{eqnarray}
\mbox{(I)}& &~ (E_1<0, E_2>0, E_3>0)\nonumber \\
& &e^{2da(t)}=\frac{1}{f^2}\left(\frac{q^2\cos^2\theta}{|l|\sinh^2[q\cos\theta
\lambda_2(t-t_2)]}
\right)^{\frac{1}{1-\alpha^2}}
\left(\frac{q^2\sin^2\theta}{r\sin^2[q\sin\theta
\lambda_3(t-t_3)]}\right)^{-\frac{\alpha^2}{1-\alpha^2}}\,,\nonumber \\
& &e^{2(D-d-2)b(t)}=\frac{f^2q^2}{|V_1|\cosh^2[q\lambda_1(t-t_1)]}
\left(\frac{q^2\cos^2\theta}{|l|\sinh^2[q\cos\theta
\lambda_2(t-t_2)]}
\right)^{-\frac{1}{1-\alpha^2}}
\left(\frac{q^2\sin^2\theta}{r\sin^2[q\sin\theta
\lambda_3(t-t_3)]}\right)^{\frac{\alpha^2}{1-\alpha^2}}\,,\nonumber \\
& &e^{2\kappa\Phi(t)}=\left(\frac{r\cos^2\theta}{|l|\sin^2\theta}
\frac{\sin^2[q\sin\theta
\lambda_3(t-t_3)]}{\sinh^2[q\cos\theta
\lambda_2(t-t_2)]}
\right)^{-\frac{\alpha}{1-\alpha^2}}
\,.
\end{eqnarray}
\begin{eqnarray}
\mbox{(II)}& &~ (E_1<0, E_2>0, E_3<0)\nonumber \\
& &e^{2da(t)}=\frac{1}{f^2}\left(\frac{q^2\cosh^2\theta}{|l|\sinh^2[q\cosh\theta
\lambda_2(t-t_2)]}
\right)^{\frac{1}{1-\alpha^2}}
\left(\frac{q^2\sinh^2\theta}{r\sinh^2[q\sinh\theta
\lambda_3(t-t_3)]}\right)^{-\frac{\alpha^2}{1-\alpha^2}}\,,\nonumber \\
& &e^{2(D-d-2)b(t)}=\frac{f^2q^2}{|V_1|\cosh^2[q\lambda_1(t-t_1)]}
\left(\frac{q^2\cosh^2\theta}{|l|\sinh^2[q\cosh\theta
\lambda_2(t-t_2)]}
\right)^{-\frac{1}{1-\alpha^2}}
\left(\frac{q^2\sinh^2\theta}{r\sinh^2[q\sinh\theta
\lambda_3(t-t_3)]}\right)^{\frac{\alpha^2}{1-\alpha^2}}\,,\nonumber \\
& &e^{2\kappa\Phi(t)}=\left(\frac{r\cosh^2\theta}{|l|\sinh^2\theta}
\frac{\sinh^2[q\sinh\theta
\lambda_3(t-t_3)]}{\sinh^2[q\cosh\theta
\lambda_2(t-t_2)]}
\right)^{-\frac{\alpha}{1-\alpha^2}}
\,.
\end{eqnarray}
\begin{eqnarray}
\mbox{(III)}& &~ (E_1<0, E_2<0, E_3>0)\nonumber \\
& &e^{2da(t)}=\frac{1}{f^2}\left(\frac{q^2\sinh^2\theta}{|l|\sin^2[q\sinh\theta
\lambda_2(t-t_2)]}
\right)^{\frac{1}{1-\alpha^2}}
\left(\frac{q^2\cosh^2\theta}{r\sin^2[q\cosh\theta
\lambda_3(t-t_3)]}\right)^{-\frac{\alpha^2}{1-\alpha^2}}\,,\nonumber \\
& &e^{2(D-d-2)b(t)}=\frac{f^2q^2}{|V_1|\cosh^2[q\lambda_1(t-t_1)]}
\left(\frac{q^2\sin^2\theta}{|l|\sin^2[q\sinh\theta
\lambda_2(t-t_2)]}
\right)^{-\frac{1}{1-\alpha^2}}
\left(\frac{q^2\cosh^2\theta}{r\sin^2[q\cosh\theta
\lambda_3(t-t_3)]}\right)^{\frac{\alpha^2}{1-\alpha^2}}\,,\nonumber \\
& &e^{2\kappa\Phi(t)}=\left(\frac{r\sinh^2\theta}{|l|\cosh^2\theta}
\frac{\sin^2[q\cosh\theta
\lambda_3(t-t_3)]}{\sin^2[q\sinh\theta
\lambda_2(t-t_2)]}
\right)^{-\frac{\alpha}{1-\alpha^2}}
\,.
\end{eqnarray}
\end{itemize}
\subsubsection{$l<0$ and $r<0$}
\begin{itemize}
\item $k_b=-1$
\begin{eqnarray}
\mbox{(I)}& &~(E_1<0, E_2>0, E_3<0)\nonumber \\
& &e^{2da(t)}=\frac{1}{f^2}\left(\frac{q^2\cosh^2\theta}{|l|\sinh^2[q\cosh\theta
\lambda_2(t-t_2)]}
\right)^{\frac{1}{1-\alpha^2}}
\left(\frac{q^2\sinh^2\theta}{|r|\cosh^2[q\sinh\theta
\lambda_3(t-t_3)]}\right)^{-\frac{\alpha^2}{1-\alpha^2}}\,,\nonumber \\
& &e^{2(D-d-2)b(t)}=\frac{f^2q^2}{V_1\sinh^2[q\lambda_1(t-t_1)]}
\left(\frac{q^2\cosh^2\theta}{|l|\sinh^2[q\cosh\theta
\lambda_2(t-t_2)]}
\right)^{-\frac{1}{1-\alpha^2}}
\left(\frac{q^2\sinh^2\theta}{|r|\cosh^2[q\sinh\theta
\lambda_3(t-t_3)]}\right)^{\frac{\alpha^2}{1-\alpha^2}}\,,\nonumber \\
& &e^{2\kappa\Phi(t)}=\left(\frac{|r|\cosh^2\theta}{|l|\sinh^2\theta}
\frac{\cosh^2[q\sinh\theta
\lambda_3(t-t_3)]}{\sinh^2[q\cosh\theta
\lambda_2(t-t_2)]}
\right)^{-\frac{\alpha}{1-\alpha^2}}
\,.
\end{eqnarray}
\begin{eqnarray}
\mbox{(II)}& &~ (E_1>0, E_2>0, E_3<0)\nonumber \\
& &e^{2da(t)}=\frac{1}{f^2}\left(\frac{q^2\sinh^2\theta}{|l|\sinh^2[q\sinh\theta
\lambda_2(t-t_2)]}
\right)^{\frac{1}{1-\alpha^2}}
\left(\frac{q^2\cosh^2\theta}{|r|\cosh^2[q\cosh\theta
\lambda_3(t-t_3)]}\right)^{-\frac{\alpha^2}{1-\alpha^2}}\,,\nonumber \\
& &e^{2(D-d-2)b(t)}=\frac{f^2q^2}{V_1\sin^2[q\lambda_1(t-t_1)]}
\left(\frac{q^2\sinh^2\theta}{|l|\sinh^2[q\sinh\theta
\lambda_2(t-t_2)]}
\right)^{-\frac{1}{1-\alpha^2}}
\left(\frac{q^2\cosh^2\theta}{|r|\cosh^2[q\cosh\theta
\lambda_3(t-t_3)]}\right)^{\frac{\alpha^2}{1-\alpha^2}}\,,\nonumber \\
& &e^{2\kappa\Phi(t)}=\left(\frac{|r|\sinh^2\theta}{|l|\cosh^2\theta}
\frac{\cosh^2[q\cosh\theta
\lambda_3(t-t_3)]}{\sinh^2[q\sinh\theta
\lambda_2(t-t_2)]}
\right)^{-\frac{\alpha}{1-\alpha^2}}
\,.
\end{eqnarray}
\begin{eqnarray}
\mbox{(III)}& &~ (E_1>0, E_2<0, E_3<0)\nonumber \\
& &e^{2da(t)}=\frac{1}{f^2}\left(\frac{q^2\cos^2\theta}{|l|\sin^2[q\cos\theta
\lambda_2(t-t_2)]}
\right)^{\frac{1}{1-\alpha^2}}
\left(\frac{q^2\sin^2\theta}{|r|\cosh^2[q\sin\theta
\lambda_3(t-t_3)]}\right)^{-\frac{\alpha^2}{1-\alpha^2}}\,,\nonumber \\
& &e^{2(D-d-2)b(t)}=\frac{f^2q^2}{V_1\sin^2[q\lambda_1(t-t_1)]}
\left(\frac{q^2\cos^2\theta}{|l|\sin^2[q\cos\theta
\lambda_2(t-t_2)]}
\right)^{-\frac{1}{1-\alpha^2}}
\left(\frac{q^2\sin^2\theta}{|r|\cosh^2[q\sin\theta
\lambda_3(t-t_3)]}\right)^{\frac{\alpha^2}{1-\alpha^2}}\,,\nonumber \\
& &e^{2\kappa\Phi(t)}=\left(\frac{|r|\cos^2\theta}{|l|\sin^2\theta}
\frac{\cosh^2[q\sin\theta
\lambda_3(t-t_3)]}{\sin^2[q\cos\theta
\lambda_2(t-t_2)]}
\right)^{-\frac{\alpha}{1-\alpha^2}}
\,.
\end{eqnarray}
\item $k_b=0$ ($E_1<0$, $E_2>0$, $E_3<0$)
\begin{eqnarray}
& &e^{2da(t)}=\frac{1}{f^2}\left(\frac{q^2\cosh^2\theta}{|l|\sinh^2[q\cosh\theta
\lambda_2(t-t_2)]}
\right)^{\frac{1}{1-\alpha^2}}
\left(\frac{q^2\sinh^2\theta}{|r|\cosh^2[q\sinh\theta
\lambda_3(t-t_3)]}\right)^{-\frac{\alpha^2}{1-\alpha^2}}\,,\nonumber \\
& &e^{2(D-d-2)b(t)}=C_1f^2e^{2q\lambda_1t}
\left(\frac{q^2\cosh^2\theta}{|l|\sinh^2[q\cosh\theta
\lambda_2(t-t_2)]}
\right)^{-\frac{1}{1-\alpha^2}}
\left(\frac{q^2\sinh^2\theta}{|r|\cosh^2[q\sinh\theta
\lambda_3(t-t_3)]}\right)^{\frac{\alpha^2}{1-\alpha^2}}\,,\nonumber \\
& &e^{2\kappa\Phi(t)}=\left(\frac{|r|\cosh^2\theta}{|l|\sinh^2\theta}
\frac{\cosh^2[q\sinh\theta
\lambda_3(t-t_3)]}{\sinh^2[q\cosh\theta
\lambda_2(t-t_2)]}
\right)^{-\frac{\alpha}{1-\alpha^2}}
\,.
\end{eqnarray}
\item $k_b=+1$ ($E_1<0$, $E_2>0$, $E_3<0$)
\begin{eqnarray}
& &e^{2da(t)}=\frac{1}{f^2}\left(\frac{q^2\cosh^2\theta}{|l|\sinh^2[q\cosh\theta
\lambda_2(t-t_2)]}
\right)^{\frac{1}{1-\alpha^2}}
\left(\frac{q^2\sinh^2\theta}{|r|\cosh^2[q\sinh\theta
\lambda_3(t-t_3)]}\right)^{-\frac{\alpha^2}{1-\alpha^2}}\,,\nonumber \\
& &e^{2(D-d-2)b(t)}=\frac{f^2q^2}{|V_1|\cosh^2[q\lambda_1(t-t_1)]}
\left(\frac{q^2\cosh^2\theta}{|l|\sinh^2[q\cosh\theta
\lambda_2(t-t_2)]}
\right)^{-\frac{1}{1-\alpha^2}}
\left(\frac{q^2\sinh^2\theta}{|r|\cosh^2[q\sinh\theta
\lambda_3(t-t_3)]}\right)^{\frac{\alpha^2}{1-\alpha^2}}\,,\nonumber \\
& &e^{2\kappa\Phi(t)}=\left(\frac{|r|\cosh^2\theta}{|l|\sinh^2\theta}
\frac{\cosh^2[q\sinh\theta
\lambda_3(t-t_3)]}{\sinh^2[q\cosh\theta
\lambda_2(t-t_2)]}
\right)^{-\frac{\alpha}{1-\alpha^2}}
\,.
\end{eqnarray}
\end{itemize}

}

\section{solutions in the model with $r=0$}
\label{r0}
In this appendix, we comment on the solutions in the model
with
$r=0$, which is substantially equivalent to the model studied in many papers
including
\cite{CGG,Roy}, and we show that the solutions are obtained as the solutions with
finite
$|r|$  by taking the small $r$ limit.

The solution for $z$ in this case is simple:
\begin{equation}
z(t)=q_3(t-t_3)\,,
\end{equation}
where  $q_3$ and $t_3$ are integration constants.
Then, the constant $E_3=\sigma\frac{q_3^2}{2}$.

When $l>0$, 
the solution for $y$ is given by
\begin{equation}
y(t)=\frac{1}{2\lambda_2}\ln \frac{q_2^2}{\cosh^2q_2\sqrt{l}f\lambda_2
(t-t_2)}\,,
\end{equation}
where $q_2$ and $t_2$ are integration constants,
and thus $E_2>0$.

Therefore, the possible cases are listed below:
\begin{tabbing}
\qquad$\sigma=+1$, $l>0$, $E_1<0$, $E_2>0$, $E_3>0$\,,\\
\qquad$\sigma=+1$, $l<0$, $E_1<0$, $E_2>0$, $E_3>0$\,,\\
\qquad$\sigma=+1$, $l<0$, $E_1<0$, $E_2<0$, $E_3>0$\,,\\
\qquad$\sigma=+1$, $l<0$, $E_1>0$, $E_2<0$, $E_3>0$\,,\\
\qquad$\sigma=-1$, $l>0$, $E_1<0$, $E_2>0$, $E_3<0$\,,\\
\qquad$\sigma=-1$, $l>0$, $E_1>0$, $E_2>0$, $E_3<0$\,,\\
\qquad$\sigma=-1$, $l<0$, $E_1<0$, $E_2>0$, $E_3<0$\,,\\
\qquad$\sigma=-1$, $l<0$, $E_1>0$, $E_2>0$, $E_3<0$\,,\\
\qquad$\sigma=-1$, $l<0$, $E_1>0$, $E_2<0$, $E_3<0$\,.
\end{tabbing}

In each case, $r=0$ can be considered as the limit $r\rightarrow 0$
of the solution for a finite $r$.
This is because if we set $|r|=e^{-2q'\lambda_3t_0}$, we find that
\begin{equation}
\lim_{q'(t-t_3-t_0)\rightarrow\infty}2|r|\cosh^2[q'\lambda_3(t-t_3)]=
\lim_{q'(t-t_3-t_0)\rightarrow\infty}2|r|\sinh^2[q'\lambda_3(t-t_3)]=
e^{2q'\lambda_3(t-t_3-t_0)}\,,
\end{equation}
noting that the time-reversal invariance of the equation of motion and the 
multiplicative constant factor is irrelevant for this case.%
\footnote{Note also that the prefactor can be absorbed into the integration
constant in the exponential function,} 
Thus, we do not exhibit the explicit form of the
solutions in the case
$r=0$ in this paper. The case $l=0$ can be analyzed similarly (and we do not
repeat the analysis). By using $l\leftrightarrow r$ symmetry, the case of
arbitrary dilaton coupling $\alpha$ can be covered by considering the limit
$r\rightarrow 0$ properly as indicated above.


\bibliographystyle{apsrev4-1}

\end{document}